\documentclass[%
 reprint,
superscriptaddress,
 amsmath,amssymb,
 aps,
]{revtex4-2}

\usepackage[colorlinks=true,allcolors=blue]{hyperref}

\usepackage{graphicx}
\graphicspath{{figures/}}
\usepackage{dcolumn}
\usepackage{bm}
\usepackage{comment}

\usepackage{xcolor}

\newcommand{\RE}[1]{#1}
\newcommand{\RT}[1]{#1}

\begin{document}

\title{Control of helix orientation in chiral magnets via lateral confinement}

\author{M. Colling \href{https://orcid.org/0009-0009-5047-817X}{\includegraphics[height=0.75em]{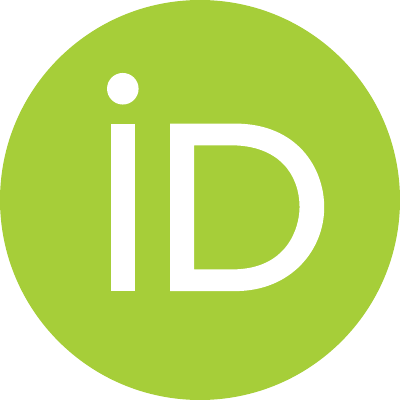}}}
\affiliation{Department of Materials Science and Engineering, Norwegian University of Science and Technology (NTNU), 7491 Trondheim, Norway}

\author{M. Stepanova \href{https://orcid.org/0000-0003-4592-4293}{\includegraphics[height=0.75em]{orcid.pdf}}} 
\affiliation{Department of Materials Science and Engineering, Norwegian University of Science and Technology (NTNU), 7491 Trondheim, Norway}

\author{M. Hentschel \href{https://orcid.org/0000-0002-6882-4183}{\includegraphics[height=0.75em]{orcid.pdf}}} 
\affiliation{4th Physics Institute and Research Center SCoPE, University of Stuttgart, 70569 Stuttgart, Germany}

\author{S. Bhattacharjee \href{https://orcid.org/0009-0007-3249-1763}{\includegraphics[height=0.75em]{orcid.pdf}}}
\affiliation{Institute of Theoretical Solid State Physics, Karlsruhe Institute of Technology, 76131 Karlsruhe, Germany}

\author{E. Lysne}
\affiliation{Department of Materials Science and Engineering, Norwegian University of Science and Technology (NTNU), 7491 Trondheim, Norway}

\author{K. Hunnestad \href{https://orcid.org/0000-0003-1732-3634}{\includegraphics[height=0.75em]{orcid.pdf}}} 
\affiliation{Department of Materials Science and Engineering, Norwegian University of Science and Technology (NTNU), 7491 Trondheim, Norway}

\author{N. Kanazawa \href{https://orcid.org/0000-0003-3270-2915}{\includegraphics[height=0.75em]{orcid.pdf}}}
\affiliation{Institute of Industrial Science, University of Tokyo, Tokyo 153-8505, Japan}

\author{Y. Tokura \href{https://orcid.org/0000-0002-2732-4983}{\includegraphics[height=0.75em]{orcid.pdf}}}
\affiliation{RIKEN Center for Emergent Matter Science (CEMS), Wako, 351-0198, Japan}
\affiliation{Department of Applied Physics, University of Tokyo, Tokyo, 113-8656, Japan}
\affiliation{Tokyo College, University of Tokyo, Tokyo 113-8656, Japan}

\author{J. Masell \href{https://orcid.org/0000-0002-9951-4452}{\includegraphics[height=0.75em]{orcid.pdf}}}
\email[Corresponding author: ]{jan.masell@kit.edu}
\affiliation{Institute of Theoretical Solid State Physics, Karlsruhe Institute of Technology, 76131 Karlsruhe, Germany}
\affiliation{RIKEN Center for Emergent Matter Science (CEMS), Wako, 351-0198, Japan}

\author{D. Meier \href{https://orcid.org/0000-0002-8623-6705}{\includegraphics[height=0.75em]{orcid.pdf}}}
\email[Corresponding author: ]{dennis.meier@uni-due.de}
\affiliation{Department of Materials Science and Engineering, Norwegian University of Science and Technology (NTNU), 7491 Trondheim, Norway}
\affiliation{Faculty of Physics and Center for Nanointegration Duisburg-Essen (CENIDE), University of Duisburg-Essen, 47057 Duisburg, Germany}
\affiliation{Research Center Future Energy Materials and Systems, Research Alliance Ruhr, 44780 Bochum, Germany}

\date{\today}

\begin{abstract}
Helimagnetic materials offer a versatile platform for spin-based device concepts owing to their long-range, tunable spiral order. 
Here, we demonstrate controlled manipulation of the helimagnetic propagation vector $\bm{q}$ by geometrical confinement, using FeGe as a model \RE{Dzyaloshinskii--Moriya interaction (DMI)-driven} chiral magnet. 
Micromagnetic simulations based on the nonlinear sigma model reveal that open boundaries give rise to a chiral surface twist acting as an effective surface anisotropy, which dictates the preferred helix orientation in the absence of magnetostatic shape effects. 
This geometry-induced anisotropy is quantitatively captured by an analytical model derived from the DMI boundary condition. 
Magnetic force microscopy measurements on focused-ion-beam structured FeGe confirm the predicted orientation behavior and establish geometry-controlled helimagnetic order as a robust, tunable mechanism for steering DMI-stabilized spin-spiral states. 
The concept provides a general route toward device-level control of chiral magnetic order in non-centrosymmetric systems.
\end{abstract}

\maketitle

\section{Introduction}
Ferromagnetic materials form an important foundation for computing technologies, most notably in magnetic memory devices such as magnetic tapes, hard disk drives, and magnetic random-access memory (MRAM)~\cite{wallace1951reproduction, chappert2007emergence, parkin2008magnetic, bhatti2017spintronics, bhushan2018historical}. 
Their clear magnetic contrast and straightforward control have made them indispensable for data storage. 
However, the strong dipolar stray fields inherent to ferromagnets cause cross-talk between neighboring elements and hinder device miniaturization, posing fundamental limitations for high-density integration and energy-efficient operation. 
To overcome these drawbacks, antiferromagnets have attracted attention for their compensated spin structure, which eliminates dipolar stray fields and enables ultrafast spin dynamics~\cite{jungwirth2016antiferromagnetic, baltz2018antiferromagnetic, jungwirth2018multiple, jungfleisch2018perspectives}. 
Yet, the same compensation makes them difficult to manipulate and detect, restricting their direct use in practical devices.
Altermagnets have recently been proposed as a symmetry-driven alternative that combines the spin polarization of ferromagnets with the magnetic compensation of antiferromagnets, potentially offering the best of both worlds~\cite{vsmejkal2022emerging, song2025altermagnets, Song2025Electrical, Yamada2025}.

\begin{figure}
    \centering
    \includegraphics[width=\columnwidth]{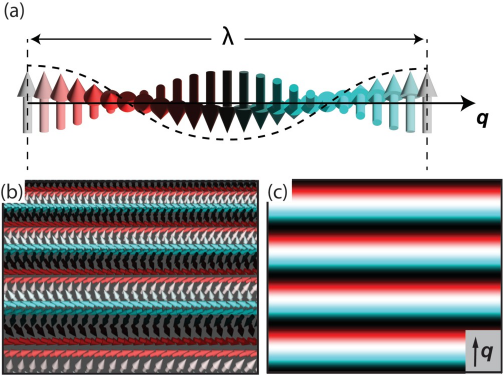}
    \caption{
    Helical spin order in FeGe. 
    (a) Schematic illustration of the helical ground state showing the wavelength $\lambda$ and propagation direction $\bm{q}$.
    (b) Micromagnetic simulation of the helical state with the magnetization shown as arrows and color-coded according to the local magnetization direction (black = down, white = up, red = right, cyan = left). 
    (c) Corresponding color-map representation of the same state as in (b).
    }
    \label{fig1}
\end{figure}

At the same time, other magnetic materials with more complex spin orders have moved into focus motivated by their unusual physical properties, which enable innovative approaches for information processing and data storage~\cite{bramwell2001spin, muhlbauer2009skyrmion, balents2010spin, xu2015discovery, gong2017discovery}. 
Among these systems, helimagnets represent an intriguing case: their noncollinear spin order combines characteristics of ferromagnets and antiferromagnets, exhibiting low or vanishing dipolar stray fields while retaining spin dynamics and transport features characteristic of ferromagnets~\cite{bogdanov1989thermodynamically, koralek2012observation, dussaux2016local, georgii2019helical, sirica2020nature}. 
Helimagnetic materials exhibit long-range spiral arrangements of spins characterized by a propagation vector $\bm{q}$, which defines the direction and periodicity of the helical order, as illustrated in Fig.~\ref{fig1}. 
Depending on the crystal symmetry and intrinsic magnetic anisotropy, $\bm{q}$ typically aligns along high-symmetry axes in the ground state~\cite{ishikawa1976helical, lebech1989magnetic}. 
However, the orientation of $\bm{q}$ can be altered by external stimuli, such as magnetic fields or electric currents, affecting key physical properties including magnetoresistance and spin-wave propagation~\cite{koralek2012observation, bauer2017symmetry, masell2020combing}. 
This tunability, which is leveraged in the emerging field referred to as helitronics, highlights the potential of helimagnetic systems for reconfigurable spin-based devices~\cite{bechler2023helitronics}.

Controlling helimagnetic order at the local length scale and in device-relevant confined geometries, however, remains a challenging task. In seminal studies, it was demonstrated that helical spin textures with $\bm{q}$ parallel to the surface emerge in manganese (Mn) monolayers on W(110)~\cite{haze2017experimental, honolka2021spin}, whereas in B20 compounds the helix axis in bulk crystals typically follows crystallographic high-symmetry directions with a tendency to align with the surface normal near exposed surfaces and in thin films~\cite{karhu2011helical, wilson2014chiral, kanazawa2016direct}. In such confined geometries, additional anisotropies become relevant: in many B20 thin films, shape anisotropy and epitaxial strain favor $\bm{q}$ along the film normal, whereas at free surfaces of bulk single crystals, surface anisotropies can stabilize an in-plane orientation of $\bm{q}$~\cite{karhu2011helical, kanazawa2016direct, dussaux2016local, schoenherr2018topological}.

This sensitivity to competing anisotropies highlights the potential for tailoring the helix orientation through geometry, interfaces, and confinement.
The examples above illustrate the general impact of geometrical confinement, which can readily be leveraged to control magnetic shape anisotropy and, hence, the propagation direction of the helimagnetic order.

In contrast to vertical confinement effects in thin-film structures, where thickness has been shown to alter the orientation and stability of helical states~\cite{wilhelm2012confinement}, the effect of lateral confinement in helimagnets is much less explored. Existing work on nanostructured chiral magnets has predominantly focused on skyrmionic textures in confined geometries, such as FeGe nanodisks~\cite{beg2015ground, beg2017dynamics}, Fe/Ir(111) islands~\cite{hagemeister2016skyrmions}, patterned ``geometric corrals''~\cite{matsumoto2022confinement}\RE{, and FeGe nanostripes~\cite{du2015edge, jin2017control}}.
While these studies primarily address skyrmions, they also reveal that the underlying helical state is strongly affected by lateral confinement --- influencing stability ranges, domain formation, and boundary behavior. 
Indeed, understanding how confinement modifies the spin helix is also crucial for controlling skyrmions as the skyrmion phase often emerges from, or coexists with, the helical phase. 

Here, we study the impact of lateral confinement on the helimagnetic order in FeGe.
Using micromagnetic simulations, we calculate the helimagnetic order in rectangular specimens of varying aspect ratio, determining the energy landscape and \RE{the} resulting $\bm{q}$-vector orientation. 
Complementary magnetic force microscopy (MFM) measurements on comparable patterns at the surface of an FeGe single crystal corroborate the simulation results, showing a substantial effect of the lateral confinement even when the helical spin textures are not fully decoupled from the magnetic order in the bulk. 
Our results give additional insights into helimagnetism in geometrically confined spaces and general guidelines for property engineering.


\section{Helical order under lateral confinement}
\label{sec2}

FeGe naturally forms a helical spin structure below $T_\mathrm{c} \approx 278\,\mathrm{K}$ with a wavelength $\lambda \approx 70\,\mathrm{nm}$~\cite{ishikawa1976helical, lebech1989magnetic}. 
The system belongs to the non-centrosymmetric B20 family and has attracted broad attention as a high-temperature skyrmion host material~\cite{yu2011near, huang2012extended, legrand2017room, leroux2018skyrmion} and a platform for studying topological dislocations and domain walls~\cite{dussaux2016local, schoenherr2018topological, schoenherr2021dislocation, stepanova2021detection}. 
Importantly for this work, its helimagnetism and fundamental physical properties are well characterized~\cite{ishikawa1976helical, yu2011near, wilhelm2012confinement, turgut2017chiral, wang2025cubic}, making FeGe an ideal model system for investigating lateral confinement effects. 
In bulk single crystals, $\bm{q}$ first aligns along $\langle 100\rangle$ crystallographic directions close to $T_\mathrm{c}$ and switches to $\langle 111\rangle$ directions upon cooling, whereas at the surface an in-plane orientation of $\bm{q}$ is generally favored \cite{lebech1989magnetic,dussaux2016local, schoenherr2018topological}. 
Aside from this surface anchoring, \RE{the in-plane orientation of $\bm{q}$ in FeGe has been shown to be nearly isotropic~\cite{dussaux2016local, schoenherr2018topological}, with the cubic magnetocrystalline anisotropy being weak near $T_\mathrm{c}$~\cite{lebech1989magnetic}. Consistent with literature~\cite{masell2020combing, bechler2023helitronics}, we therefore model the surface order observed by MFM as an isotropic two-dimensional system, where the bulk's cubic magnetocrystalline anisotropy is neglected.}

\subsection{Anisotropy due to chiral surface twist}
\label{sec2a}
To model the helimagnetic order in FeGe and calculate the energetics, we apply a similar micromagnetic model as in previous studies~\cite{rybakov2013three, beg2017dynamics} (see Supplementary Information for details). 
The model includes the standard \RE{magnetic stiffness $A_{ex}$} and bulk-type Dzyaloshinskii-Moriya interaction (DMI) \RE{$D$, see Supplementary Eq.~(S1).}
\RE{The parameters are chosen} to reproduce the experimentally observed helical wavelength of $\lambda \approx 70\,\mathrm{nm}$ and the critical field for the conical--polarized transition in bulk FeGe. 
Magnetostatic interactions are omitted \RE{because (i) the Bloch-type windings in the bulk produce no magnetic volume charges $\nabla\cdot\bm{m}$ and therefore no demagnetizing field, and (ii) the surface charges $\hat{\bm{n}}\cdot\bm{m}$, while nonzero, are negligible in the DMI-dominated regime relevant here (see Supplementary Note 2F)}.
No additional magnetic anisotropy terms are included.
Simulations are performed using an enhanced version of \textsc{MuMax3}~\cite{vansteenkiste2014design,Exl2014}, which implements higher-order finite-difference stencils to minimize artificial anisotropies originating from the discretization.
Lateral confinement is modeled by simulating rectangular specimens of varying aspect ratios with open boundary conditions, using a single discretization cell along the thickness direction to effectively represent a two-dimensional system. 
System sizes are selected to minimize commensurability artifacts between the simulated geometry and the helical wavelength.

\begin{figure}
    \centering
    \includegraphics[width=\columnwidth]{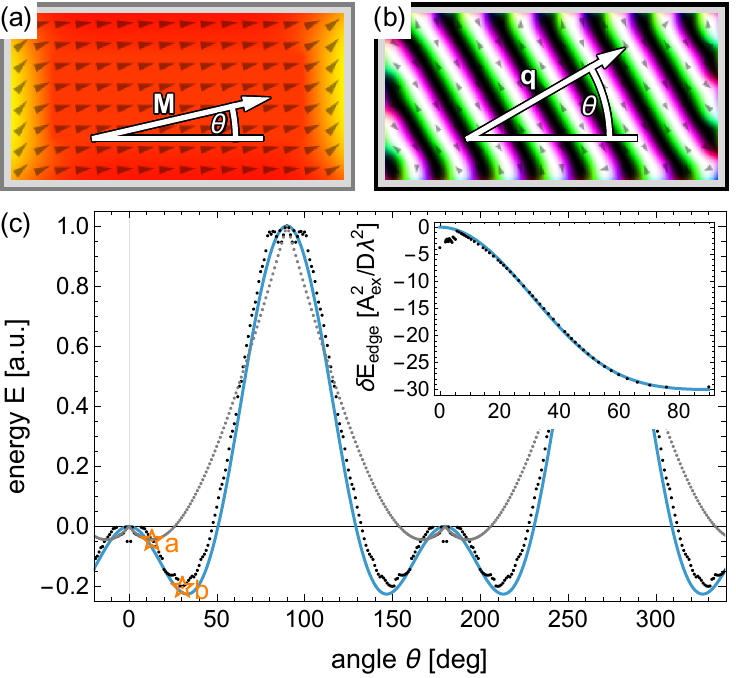}
    \caption{
    Shape anisotropy in a rectangular magnet with an aspect ratio of 2:1. 
    (a) Simulated ferromagnetic state including dipolar interactions, showing alignment of $\bm{M}$ along the long axis due to conventional shape anisotropy. 
    \RE{Color denotes the orientation of the magnetization, see Fig.~\ref{fig1}, with gray triangles emphasizing the in-plane orientation. 
    The large white arrow highlights the orientation of the magnetization in the middle of the sample and its angle $\theta$ with the x-axis.}
    (b) Simulated helical state without demagnetization energy, where the propagation vector $\bm{q}$ aligns diagonally to minimize the boundary energy through the chiral surface twist. 
    \RE{Color scheme as in panel (a), with the large white arrow indicating the orientation of the $\bm{q}$-vector in the middle of the sample.}
    (c) Corresponding energy landscapes as a function of orientation angle $\theta$.
    Orange stars mark the energy minima.
    \RT{In the main panel, the blue line is the analytical result of Eq.~(\ref{eq:Etot}), obtained by summing the surface-twist energy over the four edges of the rectangle.}
    The inset shows the calculated surface-twist contribution\RT{, with the blue line given by the analytical harmonic model of Eq.~(\ref{eq:Eedgefit})}.}
    
    \label{fig2}
\end{figure}


We first analyze magnetic anisotropy in a rectangular test structure (aspect ratio 2:1), as shown in Fig.~\ref{fig2}. Figure~\ref{fig2}a shows a ferromagnet with long-range dipolar interactions, simulated in a $560\,\mathrm{nm} \times 280\,\mathrm{nm}$ ($8\,\lambda \times 4\,\lambda$) rectangular geometry using our customized FeGe parameters, including the demagnetizing field but no DMI. During energy minimization, specific trial orientations of the magnetization $\bm{M}$ in the sample interior were fixed, while only the outer region was allowed to relax. As expected for conventional ferromagnetic order, in this configuration, the magnetization $\bm{M}$ aligns preferentially along the long side of the rectangle and remains confined to the sample plane as a direct consequence of shape anisotropy.

In Fig.~\ref{fig2}b, we show the helimagnetic order that arises when switching on the DMI but neglecting dipolar interactions. In this scenario, the associated propagation vector $\bm{q}$ aligns along the diagonal, which reduces the total edge energy by orienting the helical stripes so that both sample edges contribute equally to the boundary-induced spin rotation. This phenomenon is known as chiral surface twist~\cite{rohart2013skyrmion, meynell2014surface, muller2016edge}.
The surface twist lowers the energy for certain helix orientations relative to the boundary and thus acts as an effective surface anisotropy (see boundary condition in Eq.~(\ref{eq:BC})). The comparison of Figs.~\ref{fig2}a and \ref{fig2}b highlights a qualitatively different response to shape anisotropy: alignment of $\bm{M}$ with the long outer axis for the ferromagnet versus a diagonal orientation of $\bm{q}$ for the helimagnet.


A more detailed analysis of the two systems is displayed in Fig.~\ref{fig2}c (ferromagnet: grey dots; helimagnet: black dots), showing the energy as a function of the orientation angle $\theta$ defined in Figs.~\ref{fig2}a,b. 
For the helimagnet, the values are obtained by preparing helical states of the ideal wavelength, relaxing the edges, and optimizing the \RE{spatial} phase shift\RE{~$\phi$, which describes the translation of the helix along the $\bm{q}$-vector}. 
The residual oscillations arise from edge-related commensurability effects. 
Both systems exhibit distinct angular dependencies, with minima at $\theta = 13^\circ$ 
 (ferromagnet) and $\theta=30^\circ$ (helimagnet), marked by orange stars. 
 For the ferromagnet, the minimum is slightly offset from $\theta=0^\circ$ due to a small tilt of $\bm{M}$ near the edges. 
 By symmetry, the energy curves are mirror-symmetric about $0^\circ$ and $90^\circ$ and periodic with a periodicity of $180^\circ$.

As already noted, for chiral magnets described by Supplementary Eq.~(S1), the boundary condition\RE{~\cite{rohart2013skyrmion, meynell2014surface}}
\begin{equation}
    \partial_n\bm{m} = \frac{D}{2 A_{ex}} \, \bm{n} \times \bm{m}
    \label{eq:BC}
\end{equation}
enforces a rotation of the local magnetization $\bm{m}$ about the surface normal $\bm{n}$ on the length scale $\tfrac{2 A_{ex}}{D}=\tfrac{\lambda}{2\pi}=\tfrac{1}{|\bm{q}|}$. 
This condition is trivially fulfilled for $\bm{q}\parallel\bm{n}$, whereas for other orientations of $\bm{q}$ the energy can be lowered by additional twisting at the surface.
We numerically calculate this energy gain per surface area $\delta E_\mathrm{edge}$, which is shown in the inset of Fig.~\ref{fig2}c as black dots, displayed in dimensionless units using the energy $\tfrac{A_{ex}^2}{D}$ per area $\lambda^2$ (see Supplementary Note 2A).
\RE{The result is a smooth curve for $\theta \gtrsim 6^\circ$ with a shallow minimum around $\theta=90^\circ$. In turn, for $\theta \lesssim 6^\circ$ the projected helical wavelength diverges and the helical state becomes almost translation invariant along the edge. This invariance is broken by an energetically favorable surface state with additional twisting along the edge, referred to as stacked spirals in literature~\cite{rybakov2016new}. Real-space figures of these special states are included in Supplementary Fig.~S1.
}

Neglecting the range $\theta \lesssim 6^\circ$ where non-linear surface reconstruction sets in,
the angular dependence is well described by the lowest-order symmetry-allowed harmonics (see Supplementary Note 2A),
\begin{equation}
    \frac{\delta E_\mathrm{edge}}{\frac{A_{ex}^2}{D}/\lambda^2} = c_2 \sin^2\theta + c_4\sin^4\theta\,\,
    \RE{.}
    \label{eq:Eedgefit}
\end{equation}
Note that the zeroth order contribution (constant offset) \RE{precisely vanishes} due to the above argument that a helix with $\bm{q}\parallel \bm{n}$ could stay pristine and would not gain any energy. 
Fitting Eq.~\eqref{eq:Eedgefit} to our \RE{numerical} data \RE{in the inset of Fig.~\ref{fig2}c}, we find $c_2 \approx -55$ and $c_4 \approx 25$, which is independent of the micromagnetic constants as the problem is dimensionless.
\RE{For these parameters, the harmonic surface energy function, Eq.~\eqref{eq:Eedgefit}, is shown as the blue line in Fig.~\ref{fig2}c, matching the numerical data for $\theta \gtrsim 6^\circ$.}
\RE{Furthermore, we use the expression in Eq.~\eqref{eq:Eedgefit} to estimate the energy as function of the $\bm{q}$-vector orientation in a more complex geometry, here the rectangular sample in Fig.~\ref{fig2}b.}
The resulting function\RE{, displayed as the blue line in the main panel of Fig.~\ref{fig2}c,} reproduces the numerical data, including the energy minimum, and only minor commensurability-related oscillations are not captured.

In conclusion, the combined numerical and analytical analysis demonstrates that the chiral surface twist constitutes an effective surface anisotropy, governing the preferred helix orientation in confined geometries.

\subsection{Shape anisotropy of the helix}
\label{sec2c}
Next, we vary the width $L_x$ and height $L_y$ of our FeGe model system to investigate how the aspect ratio affects the energy landscape.

\begin{figure}[h!]
    \centering
    \includegraphics[width=\columnwidth]{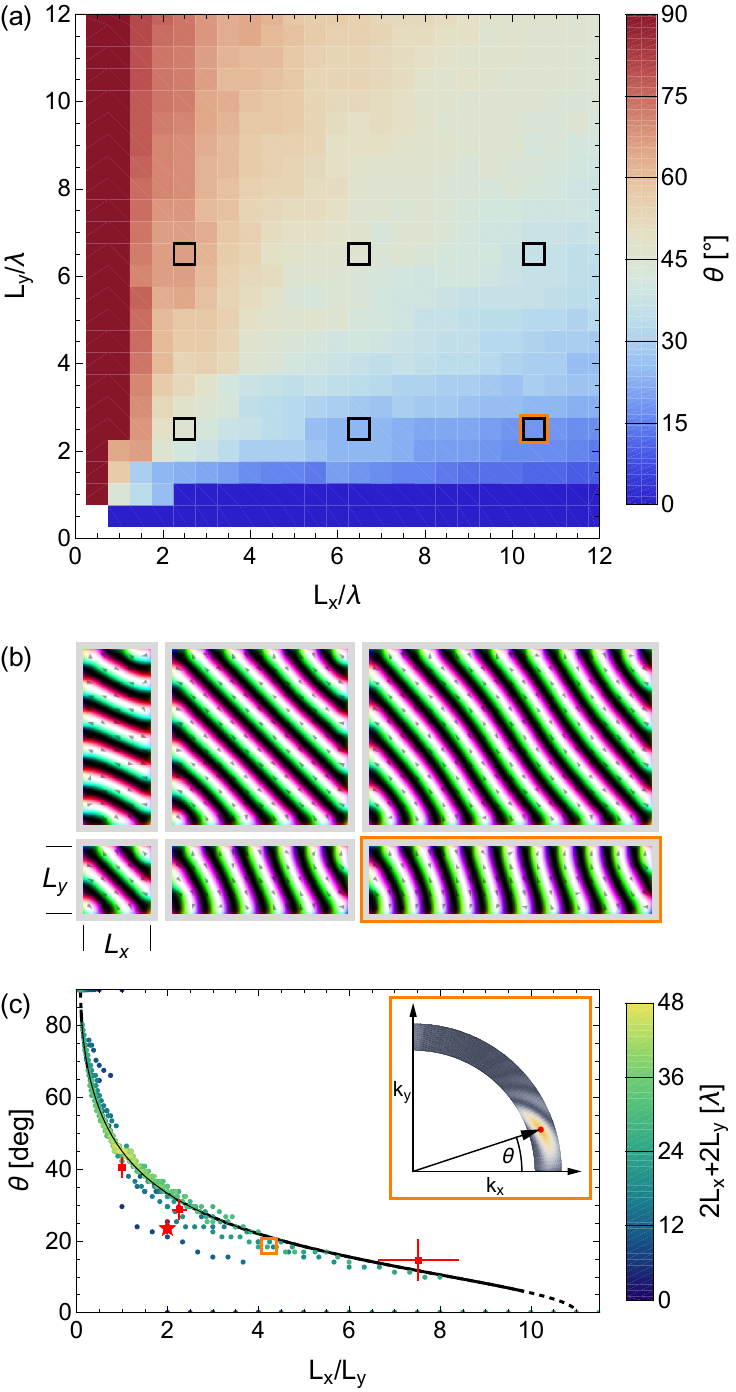}
    \caption{
    Dependence of the helix orientation on the aspect ratio of the confined region. (a) Color-coded phase diagram of the equilibrium orientation angle $\theta$ between $\bm{q}$ and $\hat{e}_x$, plotted as a function of lateral dimensions $L_x$ and $L_y$. Blue and red correspond to $\theta=0^{\circ}$ and $90^{\circ}$, respectively. Black squares mark data points for which the corresponding real-space textures are shown in (b). (b) Representative relaxed helical states for selected aspect ratios. (c) Orientation angle $\theta$ as a function of the aspect ratio $L_x/L_y$. Dots show numerical results colored by the total circumference; the black line represents the analytical model derived from the DMI boundary condition. Dashed segments denote the range outside the model’s validity ($6^{\circ} \lesssim\theta \lesssim 84^{\circ}$). The inset illustrates the restricted Fourier transform used to extract $\theta$, shown here for $(L_x,L_y) = (10.5\lambda,2.5\lambda)$, corresponding to the orange box in panels (a,b).
    }
    \label{fig3}
\end{figure}

Figure~\ref{fig3}a summarizes the energetically most favorable angle $\theta$, i.e., the angle between $\bm{q}$ and $\hat{e}_x$, as function of both $L_x$ and $L_y$. Note that in contrast to the previous section, we do not pin any parts of the magnetic texture during minimization, so that the helimagnetic texture can freely adapt its orientation to minimize the energy. Selected examples for lowest energy states are shown in Fig.~\ref{fig3}b, marked by black squares in 
Fig.~\ref{fig3}a.

We find that the phase diagram, Fig.~\ref{fig3}a, is symmetric around $L_x=L_y$, with an exception at $(1.5\lambda, 1.5\lambda)$ where the system is too small and a state with spontaneously broken symmetry emerges.
For sufficiently elongated samples, $\bm{q}$ aligns parallel to the longer axis (dark red/blue)\RE{, in agreement with previous observations on highly elongated FeGe nanostripes~\cite{du2015edge, jin2017control}}, rotating almost monotonously and smoothly as the aspect ratio varies between the two end states.
It is important to note, however, that within the sample, the helical orientation is not homogeneous and can significantly vary near the edges and corners as seen in Fig.~\ref{fig3}b.

Finally, we compare the results of the numerical minimization to the analytical model in Eq.~\eqref{eq:Eedgefit}. 
The energy of the chiral surface twist is given by
\begin{equation}
E_{\mathrm{tot}}\RE{/d} = 2L_y \,\delta E_{\mathrm{edge}}(\theta) + 2L_x \,\delta E_{\mathrm{edge}}(\theta + \pi/2)
\label{eq:Etot}
\end{equation}
\RE{where $d$ is the sample thickness.}
\RE{After some algebra, see Supplementary Note 2B, and exploiting that} $\theta$ depends only on the ratio $R=L_x/L_y$ and the ratio $c=2c_4/c_2$, \RE{we obtain the result}
\begin{equation}
\theta = \arctan \left[\sqrt{\frac{(c+1)R -1}{c-R+1}}\right]\,\,.
\end{equation}
In this expression, the fit factor $c\approx-50/55$ is obtained from fitting the data in the inset of Fig.~\ref{fig2}, as described above. 
We show the analytical result together with the full numerical data in Fig.~\ref{fig3}c. 
The agreement is good, and particularly accurate for larger systems, here quantified by a greater circumference, $2L_x+2L_y$, which are less affected by commensurability effects at the corners. 
\RE{We discuss in Supplementary Note 2E that this approach is also valid for other types of polygonal shapes.}

Figure~\ref{fig3}c also includes experimental results, indicated by a red star and red squares with error bars, which are in good agreement with the theoretical prediction. The experimental test experiments from which the data points are derived are the subject of the following section.

\section{Experimental results}
\label{sec3}


\begin{figure}
    \centering
    \includegraphics[width=\columnwidth]{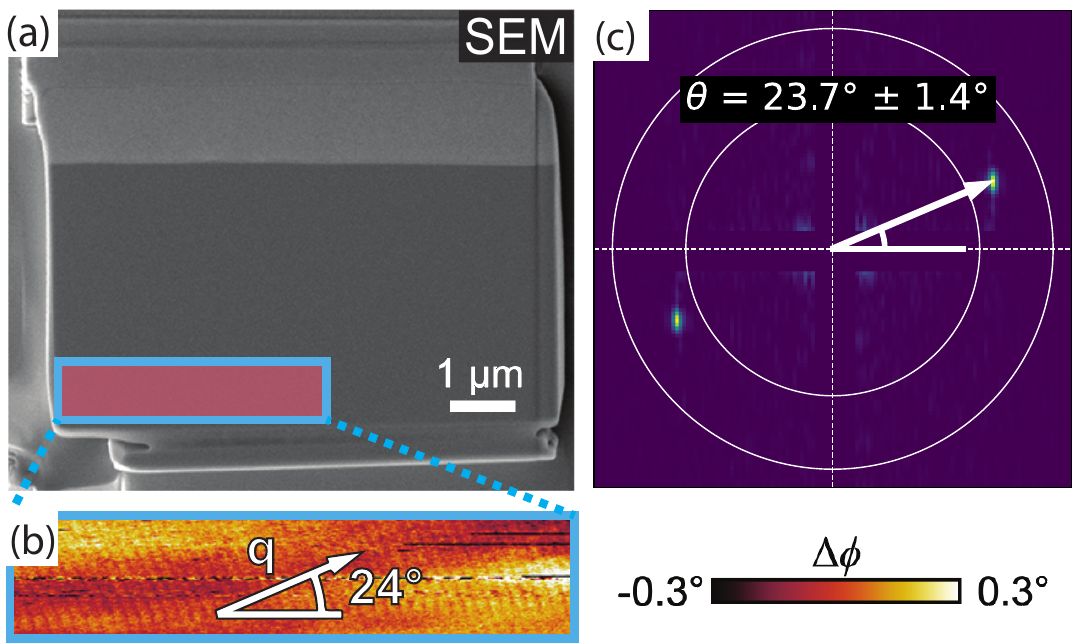}
    \caption{
    Experimental observation of helimagnetic order under lateral confinement. (a) SEM image of a 1~\textmu m thick FeGe lamella with an aspect ratio of approximately \RE{2:1}.
    (b) Corresponding MFM image revealing helimagnetic order, with propagation vector $\bm{q}$ tilted relative to the long axis. 
    (c) Fast Fourier transform (FFT) of the MFM image used to determine the orientation angle $\theta$ between $\bm{q}$ and the long axis, yielding $23.7 \pm 1.4^{\circ}$.
    }
    \label{fig4}
\end{figure}

To demonstrate the general possibility of controlling helimagnetism in laterally confined structures experimentally and verify the simulation results based on a real physical material, we conduct different experiments on FeGe lamellas and nano-structured single crystals. Figure~\ref{fig4} summarizes the results gained on a FeGe lamella that was cut from a single crystal using a focused ion beam (FIB, see Methods). A scanning electron microscopy (SEM) image of the lamella is presented in Fig.~\ref{fig4}a (thickness $\approx$ 1~$\mu$m). A representative MFM image recorded at $T = 261$~K in the area marked by the blue square in Fig.~\ref{fig4}a is displayed in Fig.~\ref{fig4}b. The image is taken in dual-pass MFM mode using a magnetized probe tip (PPP-MFMR by Nanosensors) and shows the characteristic pattern of bright and dark lines associated with the in-plane oriented spin helix in FeGe as explained in detail elsewhere~\cite{dussaux2016local}. The MFM image shows that the stripe-like pattern is not parallel to either of the sides of the lamella, exhibiting a nonzero angle $\theta$ between the propagation vector $\bm{q}$ and the positive $x$-axis (i.e., the long side of the lamella). 
Based on the Fast Fourier transformation (FFT) of the MFM signal (Fig.~\ref{fig4}c), we determine $\theta = 23.7~\pm 1.4^\circ$. This value is in reasonable agreement with the theoretically expected angle $\theta = 33^\circ$ for a sample with aspect ratio 
\RE{2:1}
, taking into account the much larger lateral dimension of the lamella for which confinement effects play a less important role than for smaller structures but cubic anisotropies of the bulk may be more relevant.
Independent of this discrepancy, this leads to two important conclusions: (i) the applied nanostructuring process by FIB preserves the helimagnetic texture and (ii) similar to the bulk single crystal from which the lamella was cut, $\bm{q}$ consistently orients within the surface plane, which is essential for its control via lateral confinement.

\begin{figure}
    \centering
    \includegraphics[width=\columnwidth]{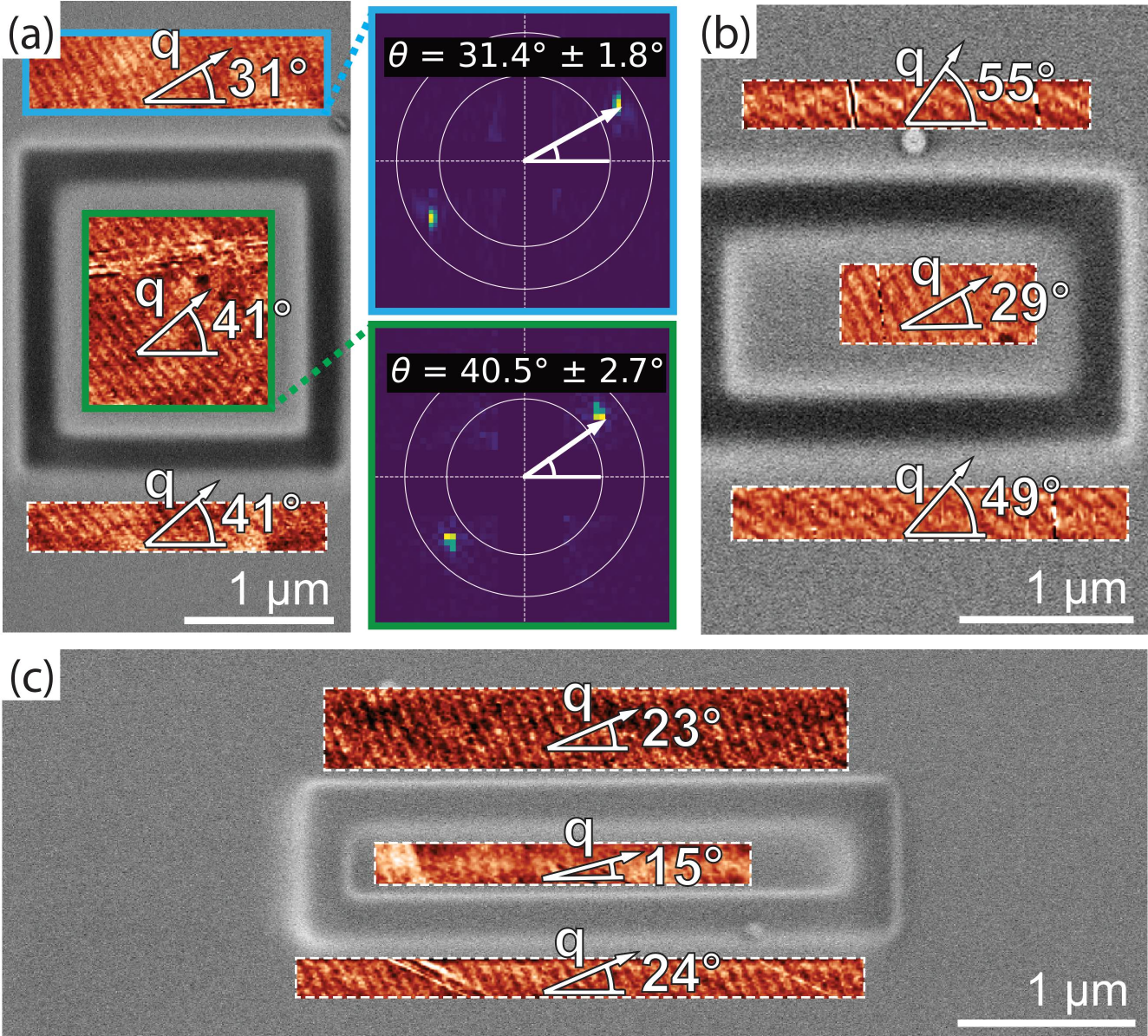}
 \caption{
    Influence of lateral confinement on the helix orientation. (a–c) SEM images of rectangular FIB-cut regions with aspect ratios of about 1:1, 2:1, and 7:1, respectively. Overlaid MFM scans show the corresponding helimagnetic stripe patterns. White arrows indicate the direction of the helical propagation vector $\bm{q}$ inside each confined region, and the extracted orientation angle $\theta$ relative to the long side of the rectangle. The comparison between interior and exterior orientations demonstrates the predictable geometry-dependent reorientation of $\bm{q}$ in agreement with simulation results.
    }
    \label{fig5}
\end{figure}

In the next step, we go to smaller length scales where confinement is expected to play a more important role and investigate the orientation of the spin helix in different rectangular regions with varying aspect ratio. For this purpose, we use FIB to cut trenches with a depth of about 150~nm
 into the surface of an FeGe single crystal, creating laterally confined regions with aspect ratios of about 1:1, 2:1, and 7:1. Although in this architecture the magnetic order in the confined structures is not decoupled from the underlying bulk, it has the advantage, compared to the lamella-based experiment, that the surface of interest has not been cut by the ion beam, reducing the risk of extrinsic defect-related pinning effects. Furthermore, right after cutting the trenches, the surface is protected by an ALD deposited Al$_2$O$_3$ capping layer (see Supplementary Fig.~S5), which protects the surface from oxidation, giving stable MFM contrast on FeGe on the timescale of months (for comparison, without capping, we observe that the MFM contrast rapidly decays and vanishes after a few days).

The experimental results are presented in Fig.~\ref{fig5}, showing overlays of SEM and MFM images recorded on different rectangles cut into the surface of the same FeGe single crystal. The FIB-cut trenches are visible in the SEM data as darker channels with a width of about 200~nm and the MFM scans reveal the orientation of $\bm{q}$ with the confined structures as well as in adjacent test regions that are displayed for comparison. To determine the local orientation of $\bm{q}$, we consider the FFT of the MFM signals, analogous to Fig.~\ref{fig4}, as exemplified for two selected regions in Fig.~\ref{fig5}a. Based on this analysis, we find the following angles $\theta$ between the propagation vector $\bm{q}$ and the positive $x$-axis, that is, $40.5 \pm 2.7^\circ$, $28.7 \pm 2.6^\circ$, and $14.6 \pm 5.6^\circ$ for aspect ratios of about 1:1, 2:1, and 7:1, respectively. 

For a direct comparison with the theoretically predicted values, the measured angles, together with their uncertainties, are plotted in red in Fig.~\ref{fig3}c, showing a remarkable agreement, taking into account that the laterally confined regions are not decoupled from the underlying bulk and its magnetic order. 
The latter has a different orientation than the helimagnetic order within the rectangles, which is confirmed by the MFM data taken in adjacent regions (see Fig.~\ref{fig5}a-c). 
Notably, in all three cases, the orientation outside the confined regions differs by up to 20$^\circ$ from the inside.
\RE{The latter reflects that the} $\bm{q}$ orientation \RE{is given by the} helimagnetic domain orientation of the bulk, which is not controlled by geometry and can vary from region to region.
Variations in the exterior angles, such as the difference between the upper ($31^\circ$) and lower ($41^\circ$) regions in Fig.~\ref{fig5}a, \RE{thus naturally occur. Importantly,} inside the confined regions, the $\bm{q}$ orientation consistently follows the predicted geometry dependence, regardless of the exterior orientation.
This observation corroborates that the lateral confinement provides a powerful handle for controlling the orientation of the helimagnetic order.

\section{Summary and conclusion}
In this work, we demonstrate that lateral confinement provides a robust and tunable handle to control the helimagnetic order in chiral magnets. 
Using FeGe as a model system, micromagnetic simulations based on the nonlinear sigma model reveal that open boundaries generate a chiral surface twist acting as an effective surface anisotropy, which governs the orientation of the helical propagation vector $\bm{q}$ in the absence of magnetostatic shape effects. 
The resulting geometry-dependent anisotropy leads to a continuous evolution of the preferred helix orientation with the sample aspect ratio, fully captured by an analytical model derived from the DMI boundary condition \RE{which is valid for rectangular geometries and beyond}.

MFM measurements on laterally confined structures cut into the surface of an FeGe single crystal confirm these predictions and demonstrate their applicability to real physical systems with the experimentally observed helix orientations inside laterally confined regions quantitatively following the simulated orientations. 
Together, these results establish geometry-induced anisotropy as a general mechanism for steering DMI-driven spin-spiral states, providing a direct route toward device-level control of chiral magnetic order. 
The concept is universal to DMI-\RE{dominated} helimagnets and can be extended to multilayer systems and synthetic chiral heterostructures where geometry and boundaries define the local anisotropy landscape.
\RE{Additional effects are expected to arise in systems with pronounced dipolar interactions (see Supplementary Fig.~S4), giving further yet-to-be-explored opportunities for controlling the helimagnetic order.}

\section{Methods}
\subsection{Micromagnetic simulations}
Micromagnetic simulations were performed using an enhanced version of \textsc{MuMax3} \RT{(version 3.10)} with higher-order finite-difference derivatives to suppress numerical anisotropies.
\RT{Verification of the higher-order stencil implementation will be published in a separate paper when the feature is made available in the next release of \textsc{MuMax3}.}
The numerical discretization is $a = \lambda/32 = 2.1875\,\textrm{nm}$.
The micromagnetic model and material parameters are given in Supplementary Note 1.

\subsection{Sample growth}
FeGe single crystals were grown by chemical vapour transport using FeGe B35 powder and I$_2$ as the transport agent. The material was sealed in an evacuated quartz ampoule and heated for one month in a three-zone furnace with a temperature gradient of 560\,$^\circ$C to 500\,$^\circ$C, yielding B20 FeGe crystals of typically $0.5 \times 1 \times 1$\,mm$^3$. The B20 structure was confirmed by powder X-ray diffraction.

\subsection{Sample preparation and nanostructuring}
Surface oxide layers were removed by argon ion beam etching (Technics Plasma R.I.B.\ Etch 160). The desired structures were then patterned into the FeGe surface using focused ion beam milling with singly charged gold ions at 35\,kV (Raith ionLine). 
After nanostructuring, the samples were coated with a 5\,nm Al$_2$O$_3$ layer deposited by ALD (Picosun R-200 Advanced) to prevent surface oxidation.

\subsection{Magnetic force microscopy measurements}
MFM measurements were performed with magnetically coated PPP-MFMR tips in two-pass mode, using the second pass to detect the magnetic signal at a lift height of approximately 30\,nm. Imaging was carried out in an NT-MDT scanning probe microscope equipped with a temperature-controlled sample holder, enabling measurements down to 260\,K. All experiments were performed in a dry nitrogen environment to ensure stable imaging conditions.

\RE{\section{Data availability}
The data and simulation scripts that support the findings of this study are openly available in the Zenodo public repository at \url{https://doi.org/10.5281/zenodo.20155860}.}

\RT{\section{Funding}
M.C. and D.M. acknowledge funding from the European Union’s Horizon Europe Programme Horizon.1.2 under the Marie Skłodowska-Curie Actions (MSCA), Grant agreement No. 101119608 (TOPOCOM).
M.S., E.L., and D.M. acknowledge funding from the Research Council of Norway, project number 263228, and support through the Norwegian Micro- and Nano-Fabrication Facility, NorFab (project number 295864).
M.S., E.L., and D.M. acknowledge support by the Research Council of Norway through its Centres of Excellence funding scheme, Project No. 262633, ``QuSpin''.
D.M. thanks NTNU for support via the Onsager Fellowship Program and the Outstanding Academic Fellows Program.
J.M. and S.B. acknowledge funding from the Deutsche Forschungsgemeinschaft (DFG, German Research Foundation) under the Project No. 547968854.}

\section{Author contributions}
M.C. wrote the manuscript and prepared Figs.~\ref{fig1}, \ref{fig4}, and \ref{fig5} under the supervision of D.M. 
J.M. wrote and tested the modified micromagnetic simulation framework.
J.M. and S.B. performed the micromagnetic simulations, analyzed the data, derived the analytical theory, and prepared Figs.~\ref{fig2} and \ref{fig3}. 
M.S. and E.L. carried out the MFM measurements. 
M.H. nanostructured and prepared the samples. 
K.H. performed the SEM imaging. 
N.K. and Y.T. provided the FeGe single crystals.
All authors discussed the results and contributed to the final version of the manuscript.

\section{Competing interests}
The authors declare no competing interests.

\bibliographystyle{naturemag} 
\bibliography{references}     

@article{wallace1951reproduction,
author={Wallace Jr., R. L.},
title={The Reproduction of Magnetically Recorded Signals},
journal={Bell System Technical Journal},
year={1951},
month={Oct},
day={01},
publisher={John Wiley {\&} Sons, Ltd},
volume={30},
number={4},
pages={1145-1173},
issn={0005-8580},
doi={10.1002/j.1538-7305.1951.tb03699.x},
url={https://doi.org/10.1002/j.1538-7305.1951.tb03699.x}
}

@article{chappert2007emergence,
author={Chappert, Claude
and Fert, Albert
and Van Dau, Fr{\'e}d{\'e}ric Nguyen},
title={The emergence of spin electronics in data storage},
journal={Nature Materials},
year={2007},
month={Nov},
day={01},
volume={6},
number={11},
pages={813-823},
issn={1476-4660},
doi={10.1038/nmat2024},
url={https://doi.org/10.1038/nmat2024}
}

@article{parkin2008magnetic,
author = {Stuart S. P. Parkin  and Masamitsu Hayashi  and Luc Thomas },
title = {Magnetic Domain-Wall Racetrack Memory},
journal = {Science},
volume = {320},
number = {5873},
pages = {190-194},
year = {2008},
doi = {10.1126/science.1145799},
URL = {https://www.science.org/doi/abs/10.1126/science.1145799}}

@article{bhatti2017spintronics,
title = {Spintronics based random access memory: a review},
journal = {Materials Today},
volume = {20},
number = {9},
pages = {530-548},
year = {2017},
issn = {1369-7021},
doi = {https://doi.org/10.1016/j.mattod.2017.07.007},
url = {https://www.sciencedirect.com/science/article/pii/S1369702117304285},
author = {Sabpreet Bhatti and Rachid Sbiaa and Atsufumi Hirohata and Hideo Ohno and Shunsuke Fukami and S.N. Piramanayagam}
}

@article{bhushan2018historical,
author={Bhushan, Bharat},
title={Historical evolution of magnetic data storage devices and related conferences},
journal={Microsystem Technologies},
year={2018},
month={Nov},
day={01},
volume={24},
number={11},
pages={4423-4436},
issn={1432-1858},
doi={10.1007/s00542-018-4133-6},
url={https://doi.org/10.1007/s00542-018-4133-6}
}

@article{jungwirth2016antiferromagnetic,
author={Jungwirth, T.
and Marti, X.
and Wadley, P.
and Wunderlich, J.},
title={Antiferromagnetic spintronics},
journal={Nature Nanotechnology},
year={2016},
month={Mar},
day={01},
volume={11},
number={3},
pages={231-241},
issn={1748-3395},
doi={10.1038/nnano.2016.18},
url={https://doi.org/10.1038/nnano.2016.18}
}

@article{baltz2018antiferromagnetic,
    title = {Antiferromagnetic spintronics},
  author = {Baltz, V. and Manchon, A. and Tsoi, M. and Moriyama, T. and Ono, T. and Tserkovnyak, Y.},
  journal = {Rev. Mod. Phys.},
  volume = {90},
  issue = {1},
  pages = {015005},
  numpages = {57},
  year = {2018},
  month = {Feb},
  publisher = {American Physical Society},
  doi = {10.1103/RevModPhys.90.015005},
  url = {https://link.aps.org/doi/10.1103/RevModPhys.90.015005}
}

@article{jungwirth2018multiple,
  author={Jungwirth, T.
and Sinova, J.
and Manchon, A.
and Marti, X.
and Wunderlich, J.
and Felser, C.},
title={The multiple directions of antiferromagnetic spintronics},
journal={Nature Physics},
year={2018},
month={Mar},
day={01},
volume={14},
number={3},
pages={200-203},
issn={1745-2481},
doi={10.1038/s41567-018-0063-6},
url={https://doi.org/10.1038/s41567-018-0063-6}
}

@article{jungfleisch2018perspectives,
  title = {Perspectives of antiferromagnetic spintronics},
journal = {Physics Letters A},
volume = {382},
number = {13},
pages = {865-871},
year = {2018},
issn = {0375-9601},
doi = {https://doi.org/10.1016/j.physleta.2018.01.008},
url = {https://www.sciencedirect.com/science/article/pii/S0375960118300343},
author = {Matthias B. Jungfleisch and Wei Zhang and Axel Hoffmann}
}

@article{vsmejkal2022emerging,
  title = {Emerging Research Landscape of Altermagnetism},
  author = {\ifmmode \check{S}\else \v{S}\fi{}mejkal, Libor and Sinova, Jairo and Jungwirth, Tomas},
  journal = {Phys. Rev. X},
  volume = {12},
  issue = {4},
  pages = {040501},
  numpages = {27},
  year = {2022},
  month = {Dec},
  publisher = {American Physical Society},
  doi = {10.1103/PhysRevX.12.040501},
  url = {https://link.aps.org/doi/10.1103/PhysRevX.12.040501}
}

@article{song2025altermagnets,
author={Song, Cheng
and Bai, Hua
and Zhou, Zhiyuan
and Han, Lei
and Reichlova, Helena
and Dil, J. Hugo
and Liu, Junwei
and Chen, Xianzhe
and Pan, Feng},
title={Altermagnets as a new class of functional materials},
journal={Nature Reviews Materials},
year={2025},
month={Jun},
day={01},
volume={10},
number={6},
pages={473-485},
issn={2058-8437},
doi={10.1038/s41578-025-00779-1},
url={https://doi.org/10.1038/s41578-025-00779-1}
}

@Article{Song2025Electrical,
author={Song, Qian
and Stavri{\'{c}}, Srdjan
and Barone, Paolo
and Droghetti, Andrea
and Antonenko, Daniil S.
and Venderbos, J{\"o}rn W. F.
and Occhialini, Connor A.
and Ilyas, Batyr
and Erge{\c{c}}en, Emre
and Gedik, Nuh
and Cheong, Sang-Wook
and Fernandes, Rafael M.
and Picozzi, Silvia
and Comin, Riccardo},
title={Electrical switching of a p-wave magnet },
journal={Nature},
year={2025},
month={Jun},
day={01},
volume={642},
number={8066},
pages={64-70},
issn={1476-4687},
doi={10.1038/s41586-025-09034-7},
url={https://doi.org/10.1038/s41586-025-09034-7}
}

@Article{Yamada2025,
author={Yamada, Rinsuke
and Birch, Max T.
and Baral, Priya R.
and Okumura, Shun
and Nakano, Ryota
and Gao, Shang
and Ezawa, Motohiko
and Nomoto, Takuya
and Masell, Jan
and Ishihara, Yuki
and Kolincio, Kamil K.
and Belopolski, Ilya
and Sagayama, Hajime
and Nakao, Hironori
and Ohishi, Kazuki
and Ohhara, Takashi
and Kiyanagi, Ryoji
and Nakajima, Taro
and Tokura, Yoshinori
and Arima, Taka-hisa
and Motome, Yukitoshi
and Hirschmann, Moritz M.
and Hirschberger, Max},
title={A metallic p-wave magnet with commensurate spin helix},
journal={Nature},
year={2025},
month={Oct},
day={01},
volume={646},
number={8086},
pages={837-842},
issn={1476-4687},
doi={10.1038/s41586-025-09633-4},
url={https://doi.org/10.1038/s41586-025-09633-4}
}

@article{bramwell2001spin,
  author = {Steven T. Bramwell  and Michel J. P. Gingras },
title = {Spin Ice State in Frustrated Magnetic Pyrochlore Materials},
journal = {Science},
volume = {294},
number = {5546},
pages = {1495-1501},
year = {2001},
doi = {10.1126/science.1064761},
URL = {https://www.science.org/doi/abs/10.1126/science.1064761}
}

@article{muhlbauer2009skyrmion,
  author = {S. Mühlbauer  and B. Binz  and F. Jonietz  and C. Pfleiderer  and A. Rosch  and A. Neubauer  and R. Georgii  and P. Böni },
title = {Skyrmion Lattice in a Chiral Magnet},
journal = {Science},
volume = {323},
number = {5916},
pages = {915-919},
year = {2009},
doi = {10.1126/science.1166767},
URL = {https://www.science.org/doi/abs/10.1126/science.1166767}
}

@article{balents2010spin,
  author={Balents, Leon},
title={Spin liquids in frustrated magnets},
journal={Nature},
year={2010},
month={Mar},
day={01},
volume={464},
number={7286},
pages={199-208},
issn={1476-4687},
doi={10.1038/nature08917},
url={https://doi.org/10.1038/nature08917}
}

@article{xu2015discovery,
  author = {Su-Yang Xu  and Ilya Belopolski  and Nasser Alidoust  and Madhab Neupane  and Guang Bian  and Chenglong Zhang  and Raman Sankar  and Guoqing Chang  and Zhujun Yuan  and Chi-Cheng Lee  and Shin-Ming Huang  and Hao Zheng  and Jie Ma  and Daniel S. Sanchez  and BaoKai Wang  and Arun Bansil  and Fangcheng Chou  and Pavel P. Shibayev  and Hsin Lin  and Shuang Jia  and M. Zahid Hasan },
title = {Discovery of a Weyl fermion semimetal and topological Fermi arcs},
journal = {Science},
volume = {349},
number = {6248},
pages = {613-617},
year = {2015},
doi = {10.1126/science.aaa9297},
URL = {https://www.science.org/doi/abs/10.1126/science.aaa9297}}

@article{gong2017discovery,
  author={Gong, Cheng
and Li, Lin
and Li, Zhenglu
and Ji, Huiwen
and Stern, Alex
and Xia, Yang
and Cao, Ting
and Bao, Wei
and Wang, Chenzhe
and Wang, Yuan
and Qiu, Z. Q.
and Cava, R. J.
and Louie, Steven G.
and Xia, Jing
and Zhang, Xiang},
title={Discovery of intrinsic ferromagnetism in two-dimensional van der Waals crystals},
journal={Nature},
year={2017},
month={Jun},
day={01},
volume={546},
number={7657},
pages={265-269},
issn={1476-4687},
doi={10.1038/nature22060},
url={https://doi.org/10.1038/nature22060}
}

@article{bogdanov1989thermodynamically,
  title={Thermodynamically stable “vortices” in magnetically ordered crystals. The mixed state of magnets},
  author={Bogdanov, Alexei N and Yablonskii, DA},
  journal={Zh. Eksp. Teor. Fiz},
  volume={95},
  number={1},
  pages={178},
  year={1989},
    url={https://api.semanticscholar.org/CorpusID:138901442}
}

@article{koralek2012observation,
  title = {Observation of Coherent Helimagnons and Gilbert Damping in an Itinerant Magnet},
  author = {Koralek, J. D. and Meier, D. and Hinton, J. P. and Bauer, A. and Parameswaran, S. A. and Vishwanath, A. and Ramesh, R. and Schoenlein, R. W. and Pfleiderer, C. and Orenstein, J.},
  journal = {Phys. Rev. Lett.},
  volume = {109},
  issue = {24},
  pages = {247204},
  numpages = {5},
  year = {2012},
  month = {Dec},
  publisher = {American Physical Society},
  doi = {10.1103/PhysRevLett.109.247204},
  url = {https://link.aps.org/doi/10.1103/PhysRevLett.109.247204}
}

@article{dussaux2016local,
  title={Local dynamics of topological magnetic defects in the itinerant helimagnet FeGe},
  author={Dussaux, Antoine and Sch{\"o}nherr, Peggy and Koumpouras, Konstantinos and Chico, Jonathan and Chang, Kevin and Lorenzelli, Luca and Kanazawa, Naoya and Tokura, Yoshinori and Garst, Markus and Bergman, Anders and others},
  journal={Nature Communications},
  volume={7},
  number={1},
  pages={12430},
  year={2016},
  publisher={Nature Publishing Group UK London},
  doi={10.1038/ncomms12430},
  url={https://doi.org/10.1038/ncomms12430}
}

@article{georgii2019helical,
AUTHOR = {Georgii, Robert and Weber, Tobias},
TITLE = {The Helical Magnet MnSi: Skyrmions and Magnons},
JOURNAL = {Quantum Beam Science},
VOLUME = {3},
YEAR = {2019},
NUMBER = {1},
ARTICLE-NUMBER = {4},
URL = {https://www.mdpi.com/2412-382X/3/1/4},
ISSN = {2412-382X},
DOI = {10.3390/qubs3010004}
}

@article{sirica2020nature,
author={Sirica, N.
and Vilmercati, P.
and Bondino, F.
and Pis, I.
and Nappini, S.
and Mo, S.-K.
and Fedorov, A. V.
and Das, P. K.
and Vobornik, I.
and Fujii, J.
and Li, L.
and Sapkota, D.
and Parker, D. S.
and Mandrus, D. G.
and Mannella, N.},
title={The nature of ferromagnetism in the chiral helimagnet Cr1/3NbS2},
journal={Communications Physics},
year={2020},
month={Apr},
day={03},
volume={3},
number={1},
pages={65},
issn={2399-3650},
doi={10.1038/s42005-020-0333-3},
url={https://doi.org/10.1038/s42005-020-0333-3}
}

@article{ishikawa1976helical,
title = {Helical spin structure in manganese silicide MnSi},
journal = {Solid State Communications},
volume = {19},
number = {6},
pages = {525-528},
year = {1976},
issn = {0038-1098},
doi = {https://doi.org/10.1016/0038-1098(76)90057-0},
url = {https://www.sciencedirect.com/science/article/pii/0038109876900570},
author = {Y. Ishikawa and K. Tajima and D. Bloch and M. Roth}
}

@article{lebech1989magnetic,
doi = {10.1088/0953-8984/1/35/010},
url = {https://doi.org/10.1088/0953-8984/1/35/010},
year = {1989},
month = {sep},
publisher = {},
volume = {1},
number = {35},
pages = {6105},
author = {B Lebech and J Bernhard and T Freltoft},
title = {Magnetic structures of cubic FeGe studied by small-angle neutron scattering},
journal = {Journal of Physics: Condensed Matter}
}

@article{bauer2017symmetry,
  title = {Symmetry breaking, slow relaxation dynamics, and topological defects at the field-induced helix reorientation in MnSi},
  author = {Bauer, A. and Chacon, A. and Wagner, M. and Halder, M. and Georgii, R. and Rosch, A. and Pfleiderer, C. and Garst, M.},
  journal = {Phys. Rev. B},
  volume = {95},
  issue = {2},
  pages = {024429},
  numpages = {12},
  year = {2017},
  month = {Jan},
  publisher = {American Physical Society},
  doi = {10.1103/PhysRevB.95.024429},
  url = {https://link.aps.org/doi/10.1103/PhysRevB.95.024429}
}

@article{masell2020combing,
  title = {Combing the helical phase of chiral magnets with electric currents},
  author = {Masell, Jan and Yu, Xiuzhen and Kanazawa, Naoya and Tokura, Yoshinori and Nagaosa, Naoto},
  journal = {Phys. Rev. B},
  volume = {102},
  issue = {18},
  pages = {180402},
  numpages = {6},
  year = {2020},
  month = {Nov},
  publisher = {American Physical Society},
  doi = {10.1103/PhysRevB.102.180402},
  url = {https://link.aps.org/doi/10.1103/PhysRevB.102.180402}
}

@article{bechler2023helitronics,
doi = {10.1088/2634-4386/ace549},
url = {https://doi.org/10.1088/2634-4386/ace549},
year = {2023},
month = {jul},
publisher = {IOP Publishing},
volume = {3},
number = {3},
pages = {034003},
author = {Bechler, N T and Masell, J},
title = {Helitronics as a potential building block for classical and unconventional computing},
journal = {Neuromorphic Computing and Engineering}
}

@article{haze2017experimental,
author={Haze, Masahiro
and Yoshida, Yasuo
and Hasegawa, Yukio},
title={Experimental verification of the rotational type of chiral spin spiral structures by spin-polarized scanning tunneling microscopy},
journal={Scientific Reports},
year={2017},
month={Oct},
day={16},
volume={7},
number={1},
pages={13269},
issn={2045-2322},
doi={10.1038/s41598-017-13329-9},
url={https://doi.org/10.1038/s41598-017-13329-9}
}

@article{honolka2021spin,
  title = {Spin-spiral state of a Mn monolayer on W(110) studied by soft x-ray absorption spectroscopy at variable temperature},
  author = {Honolka, J. and Krotzky, S. and Herzog, M. and Herden, T. and Sessi, V. and Ebert, H. and Min\'ar, J. and von Bergmann, K. and Wiesendanger, R. and \ifmmode \check{S}\else \v{S}\fi{}ipr, O.},
  journal = {Phys. Rev. B},
  volume = {103},
  issue = {17},
  pages = {174419},
  numpages = {13},
  year = {2021},
  month = {May},
  publisher = {American Physical Society},
  doi = {10.1103/PhysRevB.103.174419},
  url = {https://link.aps.org/doi/10.1103/PhysRevB.103.174419}
}

@article{karhu2011helical,
  title = {Helical magnetic order in MnSi thin films},
  author = {Karhu, E. A. and Kahwaji, S. and Robertson, M. D. and Fritzsche, H. and Kirby, B. J. and Majkrzak, C. F. and Monchesky, T. L.},
  journal = {Phys. Rev. B},
  volume = {84},
  issue = {6},
  pages = {060404},
  numpages = {4},
  year = {2011},
  month = {Aug},
  publisher = {American Physical Society},
  doi = {10.1103/PhysRevB.84.060404},
  url = {https://link.aps.org/doi/10.1103/PhysRevB.84.060404}
}

@article{wilson2014chiral,
  title = {Chiral skyrmions in cubic helimagnet films: The role of uniaxial anisotropy},
  author = {Wilson, M. N. and Butenko, A. B. and Bogdanov, A. N. and Monchesky, T. L.},
  journal = {Phys. Rev. B},
  volume = {89},
  issue = {9},
  pages = {094411},
  numpages = {13},
  year = {2014},
  month = {Mar},
  publisher = {American Physical Society},
  doi = {10.1103/PhysRevB.89.094411},
  url = {https://link.aps.org/doi/10.1103/PhysRevB.89.094411}
}

@article{kanazawa2016direct,
  title = {Direct observation of anisotropic magnetic field response of the spin helix in FeGe thin films},
  author = {Kanazawa, N. and White, J. S. and R\o{}nnow, H. M. and Dewhurst, C. D. and Fujishiro, Y. and Tsukazaki, A. and Kozuka, Y. and Kawasaki, M. and Ichikawa, M. and Kagawa, F. and Tokura, Y.},
  journal = {Phys. Rev. B},
  volume = {94},
  issue = {18},
  pages = {184432},
  numpages = {7},
  year = {2016},
  month = {Nov},
  publisher = {American Physical Society},
  doi = {10.1103/PhysRevB.94.184432},
  url = {https://link.aps.org/doi/10.1103/PhysRevB.94.184432}
}

@article{schoenherr2018topological,
author={Schoenherr, P.
and M{\"u}ller, J.
and K{\"o}hler, L.
and Rosch, A.
and Kanazawa, N.
and Tokura, Y.
and Garst, M.
and Meier, D.},
title={Topological domain walls in helimagnets},
journal={Nature Physics},
year={2018},
month={May},
day={01},
volume={14},
number={5},
pages={465-468},
issn={1745-2481},
doi={10.1038/s41567-018-0056-5},
url={https://doi.org/10.1038/s41567-018-0056-5}
}

@article{schoenherr2021dislocation,
  title={Dislocation-driven relaxation processes at the conical to helical phase transition in FeGe},
  author={Schoenherr, Peggy and Stepanova, Mariia and Lysne, Erik Nikolai and Kanazawa, Naoya and Tokura, Yoshinori and Bergman, Anders and Meier, Dennis},
  journal={ACS nano},
  volume={15},
  number={11},
  pages={17508--17514},
  year={2021},
  publisher={ACS Publications},
  doi={10.1021/acsnano.1c04302},
  url={https://doi.org/10.1021/acsnano.1c04302}
}

@article{stepanova2021detection,
  title={Detection of topological spin textures via nonlinear magnetic responses},
  author={Stepanova, Mariia and Masell, Jan and Lysne, Erik and Schoenherr, Peggy and Köhler, Laura and Paulsen, Michael and Qaiumzadeh, Alireza and Kanazawa, Naoya and Rosch, Achim and Tokura, Yoshinori and others},
  journal={Nano Letters},
  volume={22},
  number={1},
  pages={14--21},
  year={2021},
  publisher={ACS Publications},
  doi={10.1021/acs.nanolett.1c02723},
  url={https://doi.org/10.1021/acs.nanolett.1c02723}
}

@article{wilhelm2012confinement,
doi = {10.1088/0953-8984/24/29/294204},
url = {https://doi.org/10.1088/0953-8984/24/29/294204},
year = {2012},
month = {jul},
publisher = {IOP Publishing},
volume = {24},
number = {29},
pages = {294204},
author = {Wilhelm, H and Baenitz, M and Schmidt, M and Naylor, C and Lortz, R and Rößler, U K and Leonov, A A and Bogdanov, A N},
title = {Confinement of chiral magnetic modulations in the precursor region of FeGe},
journal = {Journal of Physics: Condensed Matter}
}

@article{beg2015ground,
author={Beg, Marijan
and Carey, Rebecca
and Wang, Weiwei
and Cort{\'e}s-Ortu{\~{n}}o, David
and Vousden, Mark
and Bisotti, Marc-Antonio
and Albert, Maximilian
and Chernyshenko, Dmitri
and Hovorka, Ondrej
and Stamps, Robert L.
and Fangohr, Hans},
title={Ground state search, hysteretic behaviour and reversal mechanism of skyrmionic textures in confined helimagnetic nanostructures},
journal={Scientific Reports},
year={2015},
month={Nov},
day={25},
volume={5},
number={1},
pages={17137},
issn={2045-2322},
doi={10.1038/srep17137},
url={https://doi.org/10.1038/srep17137}
}

@article{beg2017dynamics,
  title = {Dynamics of skyrmionic states in confined helimagnetic nanostructures},
  author = {Beg, Marijan and Albert, Maximilian and Bisotti, Marc-Antonio and Cort\'es-Ortu\~no, David and Wang, Weiwei and Carey, Rebecca and Vousden, Mark and Hovorka, Ondrej and Ciccarelli, Chiara and Spencer, Charles S. and Marrows, Christopher H. and Fangohr, Hans},
  journal = {Phys. Rev. B},
  volume = {95},
  issue = {1},
  pages = {014433},
  numpages = {16},
  year = {2017},
  month = {Jan},
  publisher = {American Physical Society},
  doi = {10.1103/PhysRevB.95.014433},
  url = {https://link.aps.org/doi/10.1103/PhysRevB.95.014433}
}

@article{hagemeister2016skyrmions,
  title = {Skyrmions at the Edge: Confinement Effects in $\mathrm{Fe}/\mathrm{Ir}(111)$},
  author = {Hagemeister, Julian and Iaia, Davide and Vedmedenko, Elena Y. and von Bergmann, Kirsten and Kubetzka, Andr\'e and Wiesendanger, Roland},
  journal = {Phys. Rev. Lett.},
  volume = {117},
  issue = {20},
  pages = {207202},
  numpages = {5},
  year = {2016},
  month = {Nov},
  publisher = {American Physical Society},
  doi = {10.1103/PhysRevLett.117.207202},
  url = {https://link.aps.org/doi/10.1103/PhysRevLett.117.207202}
}

@article{matsumoto2022confinement,
AUTHOR={Matsumoto, Takao  and Shibata, Naoya },
         
TITLE={Confinement of Magnetic Skyrmions to Corrals of Artificial Surface Pits with Complex Geometries},
        
JOURNAL={Frontiers in Physics},
        
VOLUME={Volume 9 - 2021},

YEAR={2022},

URL={https://www.frontiersin.org/journals/physics/articles/10.3389/fphy.2021.774951},

DOI={10.3389/fphy.2021.774951},

ISSN={2296-424X},
}

@article{du2015edge,
  title={Edge-mediated skyrmion chain and its collective dynamics in a confined geometry},
  author={Du, Haifeng and Che, Renchao and Kong, Lingyao and Zhao, Xuebing and Jin, Chiming and Wang, Chao and Yang, Jiyong and Ning, Wei and Li, Runwei and Jin, Changqing and Chen, Xianhui and Zang, Jiadong and Zhang, Yuheng and Tian, Mingliang},
  journal={Nature Communications},
  volume={6},
  pages={8504},
  year={2015},
  publisher={Nature Publishing Group},
  doi={10.1038/ncomms9504},
  url={https://doi.org/10.1038/ncomms9504}
}

@article{jin2017control,
  title={Control of morphology and formation of highly geometrically confined magnetic skyrmions},
  author={Jin, Chiming and Li, Zi-An and Kov{\'a}cs, Andr{\'a}s and Caron, Jan and Zheng, Fengshan and Rybakov, Filipp N. and Kiselev, Nikolai S. and Du, Haifeng and Bl{\"u}gel, Stefan and Tian, Mingliang and Zhang, Yuheng and Farle, Michael and Dunin-Borkowski, Rafal E.},
  journal={Nature Communications},
  volume={8},
  pages={15569},
  year={2017},
  publisher={Nature Publishing Group},
  doi={10.1038/ncomms15569},
  url={https://doi.org/10.1038/ncomms15569}
}

@article{yu2011near,
author={Yu, X. Z.
and Kanazawa, N.
and Onose, Y.
and Kimoto, K.
and Zhang, W. Z.
and Ishiwata, S.
and Matsui, Y.
and Tokura, Y.},
title={Near room-temperature formation of a skyrmion crystal in thin-films of the helimagnet FeGe},
journal={Nature Materials},
year={2011},
month={Feb},
day={01},
volume={10},
number={2},
pages={106-109},
issn={1476-4660},
doi={10.1038/nmat2916},
url={https://doi.org/10.1038/nmat2916}
}

@article{huang2012extended,
  title = {Extended Skyrmion Phase in Epitaxial $\mathrm{FeGe}(111)$ Thin Films},
  author = {Huang, S. X. and Chien, C. L.},
  journal = {Phys. Rev. Lett.},
  volume = {108},
  issue = {26},
  pages = {267201},
  numpages = {5},
  year = {2012},
  month = {Jun},
  publisher = {American Physical Society},
  doi = {10.1103/PhysRevLett.108.267201},
  url = {https://link.aps.org/doi/10.1103/PhysRevLett.108.267201}
}

@article{legrand2017room,
author = {Legrand, William and Maccariello, Davide and Reyren, Nicolas and Garcia, Karin and Moutafis, Christoforos and Moreau-Luchaire, Constance and Collin, Sophie and Bouzehouane, Karim and Cros, Vincent and Fert, Albert},
title = {Room-Temperature Current-Induced Generation and Motion of sub-100 nm Skyrmions},
journal = {Nano Letters},
volume = {17},
number = {4},
pages = {2703-2712},
year = {2017},
doi = {10.1021/acs.nanolett.7b00649},
    note ={PMID: 28358984},

URL = { https://doi.org/10.1021/acs.nanolett.7b00649
}
}

@article{leroux2018skyrmion,
author={Leroux, Maxime
and Stolt, Matthew J.
and Jin, Song
and Pete, Douglas V.
and Reichhardt, Charles
and Maiorov, Boris},
title={Skyrmion Lattice Topological Hall Effect near Room Temperature},
journal={Scientific Reports},
year={2018},
month={Oct},
day={19},
volume={8},
number={1},
pages={15510},
issn={2045-2322},
doi={10.1038/s41598-018-33560-2},
url={https://doi.org/10.1038/s41598-018-33560-2}
}

@article{turgut2017chiral,
  title = {Chiral magnetic excitations in FeGe films},
  author = {Turgut, Emrah and Park, Albert and Nguyen, Kayla and Moehle, Austin and Muller, David A. and Fuchs, Gregory D.},
  journal = {Phys. Rev. B},
  volume = {95},
  issue = {13},
  pages = {134416},
  numpages = {11},
  year = {2017},
  month = {Apr},
  publisher = {American Physical Society},
  doi = {10.1103/PhysRevB.95.134416},
  url = {https://link.aps.org/doi/10.1103/PhysRevB.95.134416}
}

@article{wang2025cubic,
author = {Wang, Kang and Wei, Wensen and Du, Haifeng},
title = {The Cubic B20 Chiral Magnet FeGe},
journal = {Advanced Functional Materials},
volume = {35},
number = {9},
pages = {2416203},
doi = {https://doi.org/10.1002/adfm.202416203},
url = {https://advanced.onlinelibrary.wiley.com/doi/abs/10.1002/adfm.202416203},
year = {2025}
}

@article{rybakov2013three,
  title = {Three-dimensional skyrmion states in thin films of cubic helimagnets},
  author = {Rybakov, F. N. and Borisov, A. B. and Bogdanov, A. N.},
  journal = {Phys. Rev. B},
  volume = {87},
  issue = {9},
  pages = {094424},
  numpages = {4},
  year = {2013},
  month = {Mar},
  publisher = {American Physical Society},
  doi = {10.1103/PhysRevB.87.094424},
  url = {https://link.aps.org/doi/10.1103/PhysRevB.87.094424}
}

@article{vansteenkiste2014design,
    author = {Vansteenkiste, Arne and Leliaert, Jonathan and Dvornik, Mykola and Helsen, Mathias and Garcia-Sanchez, Felipe and Van Waeyenberge, Bartel},
    title = {The design and verification of MuMax3},
    journal = {AIP Advances},
    volume = {4},
    number = {10},
    pages = {107133},
    year = {2014},
    month = {10},
    issn = {2158-3226},
    doi = {10.1063/1.4899186},
    url = {https://doi.org/10.1063/1.4899186},
}

@article{Exl2014,
    author  = {Exl, Lukas and
               Bance, Simon and
               Reichel, Franz and
               Schrefl, Thomas and
               {Peter Stimming}, Hans and
               Mauser, Norbert J.},
    title   = {{LaBonte's method revisited: An effective steepest
                descent method for micromagnetic energy minimization}},
    journal = {Journal of Applied Physics},
    number  = {17},
    pages   = {17D118},
    volume  = {115},
    year    = {2014},
    doi     = {10.1063/1.4862839},
    url     = {http://doi.org/10.1063/1.4862839}
}

@article{rohart2013skyrmion,
  title = {Skyrmion confinement in ultrathin film nanostructures in the presence of Dzyaloshinskii-Moriya interaction},
  author = {Rohart, S. and Thiaville, A.},
  journal = {Phys. Rev. B},
  volume = {88},
  issue = {18},
  pages = {184422},
  numpages = {8},
  year = {2013},
  month = {Nov},
  publisher = {American Physical Society},
  doi = {10.1103/PhysRevB.88.184422},
  url = {https://link.aps.org/doi/10.1103/PhysRevB.88.184422}
}

@article{meynell2014surface,
  title = {Surface twist instabilities and skyrmion states in chiral ferromagnets},
  author = {Meynell, S. A. and Wilson, M. N. and Fritzsche, H. and Bogdanov, A. N. and Monchesky, T. L.},
  journal = {Phys. Rev. B},
  volume = {90},
  issue = {1},
  pages = {014406},
  numpages = {8},
  year = {2014},
  month = {Jul},
  publisher = {American Physical Society},
  doi = {10.1103/PhysRevB.90.014406},
  url = {https://link.aps.org/doi/10.1103/PhysRevB.90.014406}
}

@article{muller2016edge,
doi = {10.1088/1367-2630/18/6/065006},
url = {https://doi.org/10.1088/1367-2630/18/6/065006},
year = {2016},
month = {jun},
publisher = {IOP Publishing},
volume = {18},
number = {6},
pages = {065006},
author = {Müller, Jan and Rosch, Achim and Garst, Markus},
title = {Edge instabilities and skyrmion creation in magnetic layers},
journal = {New Journal of Physics}
}

@article{rybakov2016new,
doi = {10.1088/1367-2630/18/4/045002},
url = {https://doi.org/10.1088/1367-2630/18/4/045002},
year = {2016},
month = {apr},
publisher = {IOP Publishing},
volume = {18},
number = {4},
pages = {045002},
author = {Rybakov, Filipp N and Borisov, Aleksandr B and Blügel, Stefan and Kiselev, Nikolai S},
title = {New spiral state and skyrmion lattice in 3D model of chiral magnets},
journal = {New Journal of Physics}
}

\end{document}


\title{Supplementary Information: Control of helix orientation in chiral magnets via lateral confinement}

\author{M. Colling \href{https://orcid.org/0009-0009-5047-817X}{\includegraphics[height=0.75em]{orcid.pdf}}}
\affiliation{Department of Materials Science and Engineering, Norwegian University of Science and Technology (NTNU), 7491 Trondheim, Norway}

\author{M. Stepanova \href{https://orcid.org/0000-0003-4592-4293}{\includegraphics[height=0.75em]{orcid.pdf}}} 
\affiliation{Department of Materials Science and Engineering, Norwegian University of Science and Technology (NTNU), 7491 Trondheim, Norway}

\author{M. Hentschel \href{https://orcid.org/0000-0002-6882-4183}{\includegraphics[height=0.75em]{orcid.pdf}}} 
\affiliation{Physics Institute and Research Center SCoPE, University of Stuttgart, Stuttgart, Germany}

\author{S. Bhattacharjee \href{https://orcid.org/0009-0007-3249-1763}{\includegraphics[height=0.75em]{orcid.pdf}}}
\affiliation{Institute of Theoretical Solid State Physics, Karlsruhe Institute of Technology, 76131 Karlsruhe, Germany}

\author{E. Lysne}
\affiliation{Department of Materials Science and Engineering, Norwegian University of Science and Technology (NTNU), 7491 Trondheim, Norway}

\author{K. Hunnestad \href{https://orcid.org/0000-0003-1732-3634}{\includegraphics[height=0.75em]{orcid.pdf}}} 
\affiliation{Department of Materials Science and Engineering, Norwegian University of Science and Technology (NTNU), 7491 Trondheim, Norway}

\author{N. Kanazawa \href{https://orcid.org/0000-0003-3270-2915}{\includegraphics[height=0.75em]{orcid.pdf}}}
\affiliation{Institute of Industrial Science, University of Tokyo, Tokyo 153-8505, Japan}

\author{Y. Tokura \href{https://orcid.org/0000-0002-2732-4983}{\includegraphics[height=0.75em]{orcid.pdf}}}
\affiliation{RIKEN Center for Emergent Matter Science (CEMS), Wako, 351-0198, Japan}
\affiliation{Department of Applied Physics, University of Tokyo, Tokyo, 113-8656, Japan}
\affiliation{Tokyo College, University of Tokyo, Tokyo 113-8656, Japan}

\author{J. Masell \href{https://orcid.org/0000-0002-9951-4452}{\includegraphics[height=0.75em]{orcid.pdf}}}
\email{jan.masell@kit.edu}
\affiliation{Institute of Theoretical Solid State Physics, Karlsruhe Institute of Technology, 76131 Karlsruhe, Germany}
\affiliation{RIKEN Center for Emergent Matter Science (CEMS), Wako, 351-0198, Japan}

\author{D. Meier \href{https://orcid.org/0000-0002-8623-6705}{\includegraphics[height=0.75em]{orcid.pdf}}} 
\email{dennis.meier@uni-due.de}
\affiliation{Department of Materials Science and Engineering, Norwegian University of Science and Technology (NTNU), 7491 Trondheim, Norway}
\affiliation{Faculty of Physics and Center for Nanointegration Duisburg-Essen (CENIDE), University of Duisburg-Essen, Duisburg, Germany}
\affiliation{Research Center Future Energy Materials and Systems, Research Alliance Ruhr, 44780 Bochum, Germany}

\maketitle

\section*{Supplementary Information}

\noindent
This Supplementary Information provides additional details on the micromagnetic model, numerical implementation, and experimental preparation procedures discussed in the main text. \ref{sec:S1} describes the micromagnetic model and parameterization used for FeGe, \ref{sec:S2} outlines the numerical implementation and data analysis procedures, and \ref{sec:S3} describes sample preparation for MFM measurements.

\section{Micromagnetic Model}
\label{sec:S1}

For completeness, we recall that bulk FeGe exhibits a Bloch-type helical order with a wavelength of $\lambda = 70\,\mathrm{nm}$~\cite{ishikawa1976helical, lebech1989magnetic}.
All simulations were performed well below the ordering temperature, assuming a constant magnetization amplitude of $M_s = 175\,\mathrm{kA/m}$ at $T = 260\,\mathrm{K}$.

Since the net magnetization in the helical phase vanishes and Bloch-type windings do not produce magnetic volume charges, the magnetostatic (demagnetizing) field is omitted.
We restrict our considerations to an effectively two-dimensional magnet of thickness $d$ in rectangular geometries, which is sufficient to capture the influence of sample shape on the helix orientation.
The normalized magnetization $\bm{m} = \bm{M}/M_s$ is described, in lowest-order gradient expansion, by the nonlinear sigma model:
\begin{equation}
    \frac{E}{d} = \int \left[
        A_{\mathrm{ex}} (\nabla\bm{m})^2
        + D\,\bm{m}\cdot(\nabla\times \bm{m})
    \right]\, d^2r .
    \label{eq:SEnergy}
\end{equation}

For FeGe, we used a magnetic stiffness of $A_{\mathrm{ex}} = 0.868\,\mathrm{pJ/m}$ and a Dzyaloshinskii--Moriya interaction constant of $D = 0.156\,\mathrm{mJ/m^2}$.
These values differ slightly from some reports in the literature, as they are derived directly from the measured helical wavelength $\lambda = 4\pi A_{\mathrm{ex}}/D$ together with the critical field for the conical-to-polarized phase transition, $\mu_0 H_{c2} = D^2/(2 A_{\mathrm{ex}} M_s)$, corrected for demagnetizing effects using data from FeGe single crystals at $T=260\,\mathrm{K}$.

This simplified model has only two intrinsic parameters, $A_{\mathrm{ex}}$ and $D$, which set the length scale $\lambda$ and an overall energy scale.
Therefore, lengths  are expressed in units of $\lambda$, and energies and energy densities are expressed in arbitrary units, making the results independent of specific material parameters.
Additional anisotropy terms are not included.
Although cubic anisotropies can influence the helix orientation in chiral magnets~\cite{bauer2017symmetry}, they are intentionally excluded here to isolate the effect of geometry-induced anisotropy.

\section{Numerical Implementation}\label{sec:S2}

Simulations were performed using an enhanced version of \textsc{MuMax3}~\cite{vansteenkiste2014design,Exl2014},
in which the finite-difference derivatives in the bulk are computed via standard, symmetric five-point stencils which yield $\mathcal{O}(a^4)$ accuracy.
At the edges of the material we impose the boundary conditions by use of asymmetric stencils.
We have verified that the error on the magnetic textures converges with $\mathcal{O}(a^4)$ in the bulk.
This higher-order scheme suppresses the parasitic cubic anisotropies that appear when using the default implementation $\mathcal{O}(a^2)$.

To be more precise, using standard \textsc{MuMax3} leads to an unphysical, numerical artifact, which is an anisotropy for the $\vec{q}$-direction, which in 2d favors alignment in the $[110]$ direction.
This anisotropy is as large as $\delta E_{num}\approx17 \tfrac{A_{ex}^2}{D}/\lambda^3$ for a helix which is discretized on $12$ lattice sites.
For finer discretization $a$, this numerical error scales with $\mathcal{O}(a^2)$ so that one obtains $\delta E_{num}\approx9.4 \tfrac{A_{ex}^2}{D}/\lambda^3$ for $\lambda/a=16$ and $\delta E_{num}\approx2.4 \tfrac{A_{ex}^2}{D}/\lambda^3$ for $\lambda/a=32$.
We see that for a system with a length of only a few $\lambda$, the energy of the unphysical anisotropy $\delta E_{num}$ dominates over the energy gain of the chiral surface twist which is of the order $\delta E_{edge} \sim 10 \tfrac{A_{ex}^2}{D}/\lambda^2$.
However, switching to our five-point-stencils with $\mathcal{O}(a^4)$ error scaling, the numerical error can be much faster suppressed, leading to $\delta E_{num}\approx0.5 \tfrac{A_{ex}^2}{D}/\lambda^3$ for $\lambda/a=16$ or $\delta E_{num}\approx0.03 \tfrac{A_{ex}^2}{D}/\lambda^3$ for $\lambda/a=32$ at no significant additional runtime of the simulations.
Note that even finer discretizations do not necessarily lead to more precise results as \textsc{MuMax3} runs on GPUs with floating point (single) precision only, which imposes another accuracy limit if differences between neighboring sites become too small or samples become too large.

Therefore, we chose a cell size of $a = \lambda/32$ throughout this manuscript in order to efficiently remove numerical anisotropy artifacts.
In all simulations, the sample thickness was $d = a$, corresponding to a single discretization cell along the $z$-direction, effectively modeling a monolayer.

\subsection{Exploiting symmetry}
\label{sec:symmetry}
The rectangular shapes have a $180^{\circ}$ rotation symmetry around the z-axis as well as a mirror symmetry with respect to the major axes (x-axis or y-axis).
Correspondingly, the energy $E(\theta)$ of a configuration described by the orientation angle $\theta$ also obeys these symmetries.
The energy is symmetric both around $\theta=0^{\circ}$ and $\theta=90^{\circ}$, and, moreover, periodic under $180^{\circ}$ rotations.
We exploited these symmetries to save on computational resources and reconstructed missing data from these symmetries.
Moreover, this leads to the lowest order harmonics description
\begin{equation}
    \delta E_\mathrm{edge} = \frac{A_{ex}}{D \lambda^2} \left(e_0 + e_2 \cos2\theta + e_4\cos4\theta\right)
    \,\,,
    \label{eq:SSurf}
\end{equation}
which applies in general both to the rectangular samples and the half-infinite samples used to compute the energy of the chiral surface twist.
This formulation is equivalent to the form used in the main text\RE{:
\begin{equation}
    \frac{\delta E_\mathrm{edge}}{\frac{A_{ex}^2}{D}/\lambda^2} = c_2 \sin^2\theta + c_4\sin^4\theta\,\,.
    \label{eq:SEedgefit}
\end{equation}
}

\RE{
\subsection{Derivation of Eq.~4: $\theta_\mathrm{min}(L_x,L_y)$}

The total energy due to the chiral surface twist in a rectangular sample having side lengths $L_x$ and $L_y$ and thickness $D$ can be expressed via the edge energy $\delta E_{\mathrm{edge}}(\theta)$, given in Eq.~(2) in the main text.
Here, $\theta$ describes the angle between $\vec{q}$ and the surface normal $\vec{n}$.
Thus, for example, the surface energy density of the bottom edge with length $L_x$ is given by  $\delta E_{\mathrm{edge}}(\theta + \pi/2)$.
The total energy, as given by Eq.~(3) in the main text, is:
\begin{equation}
\begin{split}
   E_{\mathrm{tot}}(\theta)/d &= 2L_y \,\delta E_{\mathrm{edge}}(\theta) + 2L_x \,\delta E_{\mathrm{edge}}(\theta + \pi/2) \\
   &= 2 \Big[ c_2 (L_y \, \sin^2\theta + L_x \, \cos^2\theta ) \\
   &\qquad\quad+ c_4(L_y \, \sin^4\theta + L_x \, \cos^4\theta) \Big]     
\end{split}
\end{equation}
Minimizing $E_{\mathrm{tot}}(\theta)$ with respect to $\theta$ we get
\begin{equation}
\begin{split}
   0 &=  2 \Big[ c_2 (L_y \, 2\sin\theta \cos\theta - L_x \, 2\sin\theta\cos\theta )\\
   & \qquad\quad+ c_4(L_y \, 4\sin^3\theta \cos\theta - L_x \, 4 \sin\theta\cos^3\theta) \Big] 
\end{split}
\end{equation}
and thus
\begin{equation}
   0= 4\sin\theta \cos\theta \Big[ c_2 (L_y  - L_x) + 2c_4(L_y \sin^2\theta - L_x \cos^2\theta) \Big] .
\end{equation}
We obtain the general solution from the last component after some algebra:
\begin{equation}
   \tan^2\theta = \frac{2 c_4 L_x - c_2L_y + c_2 L_x}{c_2L_y - c_2 L_x + 2 c_4 L_y}
\end{equation}
Introducing the ratio of potential constants, $c=\frac{2c_4}{c_2}$, and the ratio of side lengths, $R=\frac{L_x}{L_y}$, we obtain the final result
\begin{equation}
   \theta_\mathrm{min} = \arctan \left( \sqrt{\frac{(c+1)R -1}{c-R+1}} \right)
\end{equation}
which appears as Eq.~(4) in the main text.
}

\subsection{Numerical Analysis for Fig.~2}

\subsubsection*{Ferromagnet with Dipolar Interactions (Fig.~2a)}
A uniformly polarized configuration with a given in-plane orientation of the magnetization $\bm{M}$ was first scripted, after which the energy of an outer region of thickness $\lambda$ near the edges of the sample was minimized, while the magnetization in the interior of the sample was kept fixed.
This procedure was repeated for several orientations of $\bm{M}$ and for a variety of symmetry-distinct textures in the outer region.

The energy dependence on the orientation angle $\theta$ is shown in Fig.~2c (gray dots).
The lowest-energy state occurs at $\theta = 13^{\circ}$ due to small edge-induced tilts of $\vec{m}$.

\subsubsection*{Helimagnet without Dipolar Interactions (Fig.~2b--c)}
Helical textures with the ideal wavelength $\lambda$ were initialized for a range of propagation-vector orientations $\bm{q}$ and subsequently relaxed while keeping only the outermost region ($\sim\lambda$) free.
For each angle $\theta$, the phase shift $\phi$ was optimized by minimizing the energy independently for different values $\phi = 0^{\circ}, 1^{\circ}, \dots, 180^{\circ}$.
The resulting energies $E(\theta)$ are plotted as black dots in Fig.~2c.
Residual oscillations are in fact not noise but proper features which originate from edge-related commensurability effects between the helical stripe period and sample boundaries.

\subsubsection*{Energy gain of the chiral surface twist (Fig.~2c)}
\label{sec:chiralsurfacetwist}

\begin{figure}[t]
    \centering
    \includegraphics[width=\columnwidth]{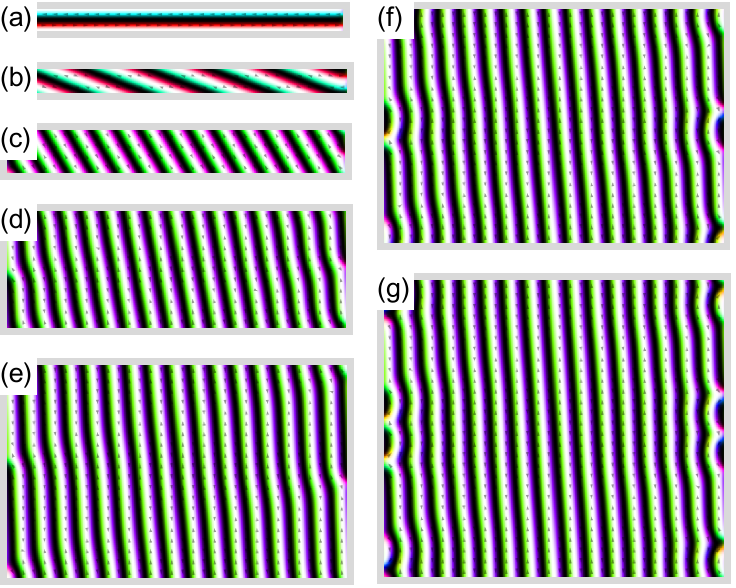}
    \caption{
        Textures for the calculation of the chiral surface twist energy.
        Panels show the relaxed states, started from a scripted pristine helix, which is commensurable with the periodic boundary conditions in the y-direction.
        The angles are (a) $\theta=90^\circ$, (b) $\theta\approx70^\circ$, (c) $\theta=30^\circ$, (d) $\theta\approx11.5^\circ$, (e) $\theta\approx5.7^\circ$, (f) $\theta\approx5.2^\circ$, and (g) $\theta\approx4.1^\circ$.
    }
    \label{fig:S1}
\end{figure}

The data in the inset of Fig.~2c shows the energy gained by the surface twist as function of the angle between $\vec{q}$ and the normal of the surface.
To generate the data, a rectangular sample of length $16\lambda$ in the $x$-direction (512 lattice sites) was simulated with open boundary conditions, Eq.~(1) of the main text.
Along the $y$-direction, the size $N_y$ was varied under \emph{periodic} boundary conditions to ensure commensurability of tilted helices.
We verified that constraining the central two lattice sites along $x$ during relaxation did not affect the results.
A selection of textures after relaxation of the energy, Eq.~\eqref{eq:SEnergy}, is shown in Fig.~\ref{fig:S1}.
Notably, the non-linear behavior for $\theta\lesssim5.5^\circ$ is clearly visible as almost translation invariant texture at the surface spontaneously forms additional waves.

\subsection{Numerical Analysis for Fig.~3}
\label{sec:2e}

The phase diagram in Fig.~3a was obtained using numerical minimizations of the micromagnetic energy Eq.~\eqref{eq:SEnergy} under open boundary conditions.
For each pair of lateral dimensions $(L_x,L_y)$ the ground state was identified by performing multiple minimizations from different initial conditions.
Initial states were scripted as helices with the ideal wavelength $|{\bm q}| = D/(2A_{\mathrm{ex}})$, prepared for a range of orientations $\theta = 0^{\circ}, 5^{\circ}, \dots, 90^{\circ}$ and phase shifts $\phi = 0^{\circ}, 11.25^{\circ}, 22.5^{\circ}, \dots, 90^{\circ}$.
This sampling covers all monodomain helical states compatible with a rectangular geometry.
To avoid numerical artifacts, these large-scale simulations used $a = \lambda/32$ to reproduce the overall phase diagram.

The lowest-energy configuration from this ensemble was selected as the ground state.
The final $\bm{q}$-orientation was determined from the peak position in a restricted Fourier transform of the relaxed texture.
An example of this Fourier analysis is shown in the inset of Fig.~3c for $(L_x,L_y) = (10.5\lambda,2.5\lambda)$, corresponding to the orange-marked data point in Fig.~3.

\RE{
\subsection{Polygonal samples beyond quadrangles}
Going beyond square-shaped or rectangular samples of different aspect ratio, we have explored the effect of lateral confinement in other polygonal samples and checked the validity of our analytical theory. 
To this end, we have considered two different types of triangles which are discussed in the following.

\subsubsection*{Equilateral triangle}
\begin{figure}[t]
    \centering
    \includegraphics[width=\columnwidth]{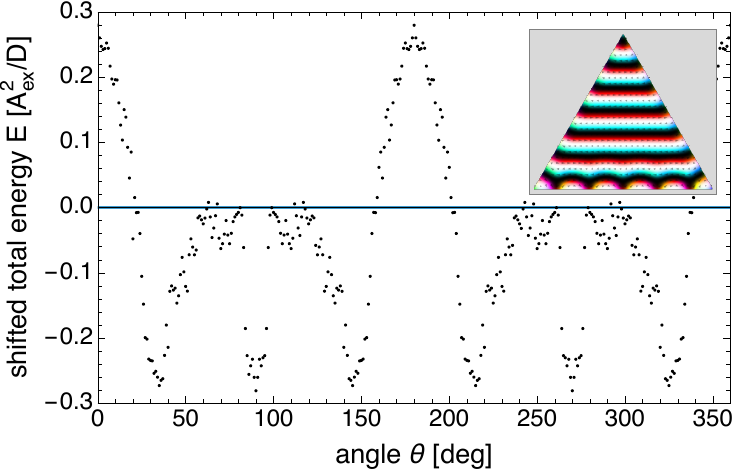}
    \caption{\RE{Energy as function of the $\bm{q}$-vector orientation in an equilateral triangle with an edge length of $554.26nm \approx 7.92 \lambda$.
    The black dots are numerical data and the solid blue line is the result obtained from analytical model. The inset shows the relaxed helical texture corresponding to $\theta=90^\circ$ (one of the minima).
    }}
    \label{fig:S2}
\end{figure}
From a $8\lambda \times8\lambda$ geometry by subtracting three rotated and translated rectangles accordingly, we get an equilateral triangle of sides $7.92\lambda$. 
The base of the triangle is smooth, but the other two sides are pixelated. 
Fig.~\ref{fig:S2} shows the energy landscape as a function of the orientation angle of $\vec{q}$-vector. 
The black dots are numerical data, exhibiting a periodicity of $180^\circ$ and mirror symmetries about $0^\circ$ and $90^\circ$. 
We lost the mirror symmetries about the other two minima at $35^\circ$ and $145^\circ$ for having two pixelated sides in the triangle. 
Thus the profile does not reflect the $3$-fold rotational symmetry of the equilateral triangle.

In our analytical model, we have considered smooth sides of the triangle and it yields a constant profile showed using blue line in Fig.~\ref{fig:S2}. 
This is expected as the energy described by Eq.~(\ref{eq:SEedgefit}) is approximated up to fourth order harmonics. 
Since the $3$-fold symmetric triangle is incommensurate with both the second and fourth order harmonics, their contributions to the angular dependence of the helix must vanish and only harmonics starting from the sixth order contribute.

\subsubsection*{The Right Triangle}
Our motivation behind exploring the right-angled triangle of sides $6\lambda$, $8\lambda$ and $10\lambda$ is that it is asymmetric, without any nontrivial mirror or rotational symmetry. 
Consequently, the second and fourth order harmonics, unlike in the case of the equilateral triangle, contribute substantially and Fig~\ref{fig:S3} shows excellent agreement of the analytical model with the numerical data. 
\begin{figure}[t]
    \centering
    \includegraphics[width=\columnwidth]{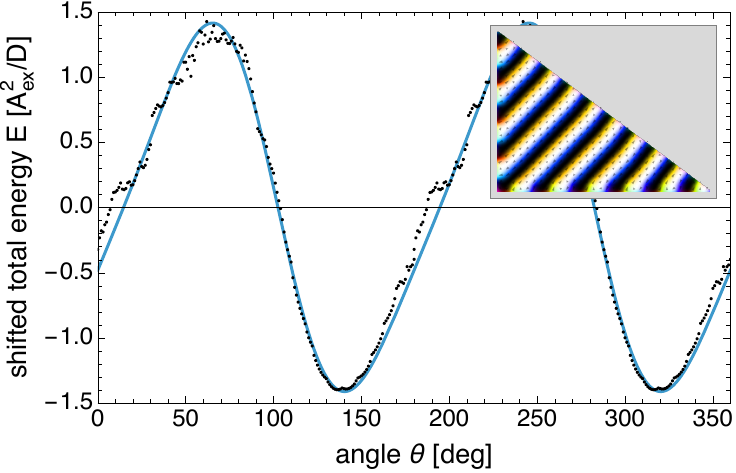}
    \caption{\RE{Energy profile as a function of the orientation angle $\theta$ for a right-angled triangle with opposite, adjacent and hypotenuse of $6\lambda = 420nm$, $8\lambda = 560nm$ and $10\lambda = 700nm$ respectively. The inset image is the relaxed helical texture of the energy minimum at $\theta = 136^\circ$.}}
    \label{fig:S3}
\end{figure}
}

\RE{
\subsection{Effect of dipolar interactions}

We have carefully analyzed the impact of dipolar interactions on the helical $\vec{q}$-vector orientation to make sure that this long-range interaction, which famously leads to shape anisotropy in ferromagnets, is irrelevant in the helimagnetic case.

As a test system, we have chosen the 2:1 geometry from Fig.~2 of the main text. 
We have analyzed the orientation dependence of a helical state as outlined in \ref{sec:2e}, with $\theta$ in increments of $5^\circ$ and the phase shifts $\phi$ in steps of $10^\circ$.
In the case of dipolar interaction, the result depends on the thickness of the sample. 
Here, we increased the thickness with additional periodic boundary conditions in the z-direction until the results converged, for which $L_z>4\lambda$ was found to be sufficient for the present case with $L_x=8\lambda$ and $L_y=4\lambda$.

\begin{figure}[t]
    \centering
    \includegraphics[width=\columnwidth]{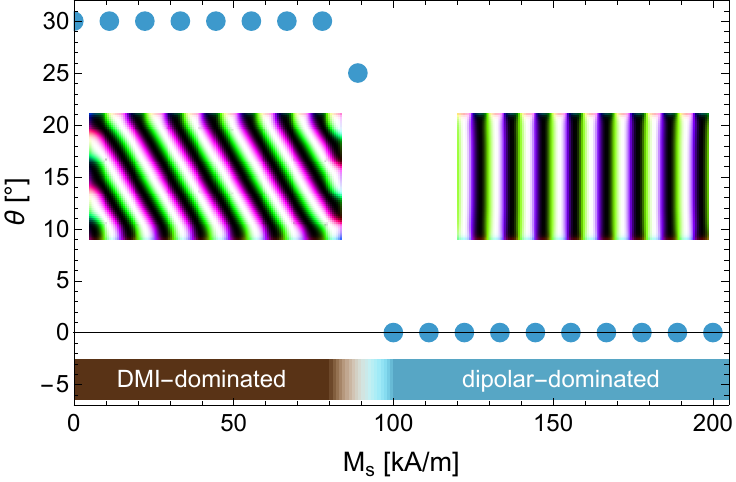}
    \caption{
        \RE{
        Helical spin order in FeGe with dipolar interaction.
        The main panel shows the orientation of the $\bm{q}$-vector as function of the saturation magnetization $M_s$.
        Insets show the real-space magnetization of the helical state for the lowest and highest value of $M_s$, respectively.
        }
    }
    \label{fig:R1}
\end{figure}

The result is shown in Fig.~\ref{fig:R1}.
Two regimes are clearly distinguishable: The DMI-dominated regime and the dipolar interaction-dominated regime.
We also computed a sparse version of the phase diagram in Fig.~3 of the main text to confirm that the helix orientation behavior in the dipolar dominated scenario is completely different from the DMI-dominated scenario, namely always shows a perfect polarization of the $\vec{q}$-vector along the longest sample axis.
While the literature values of the material parameters suggest that we should be deep in the dipolar-dominated regime, the numerical results disagree completely with our experimental observations.
In contrast,  the DMI-dominated regime precisely describes the experimental data.
We conclude that the material parameters in our case deviate from literature values and our sample is in the DMI-dominated regime.

As the two regimes are sharply separated, we conclude that dipolar interactions do not need to be considered for a sample like ours that is in the DMI-dominated regime.
The precise analysis of the two regimes and their transition will be subject of a future study.

}

\section{Sample preparation}
\label{sec:S3}

\begin{figure}[t]
    \centering
    \includegraphics[width=\columnwidth]{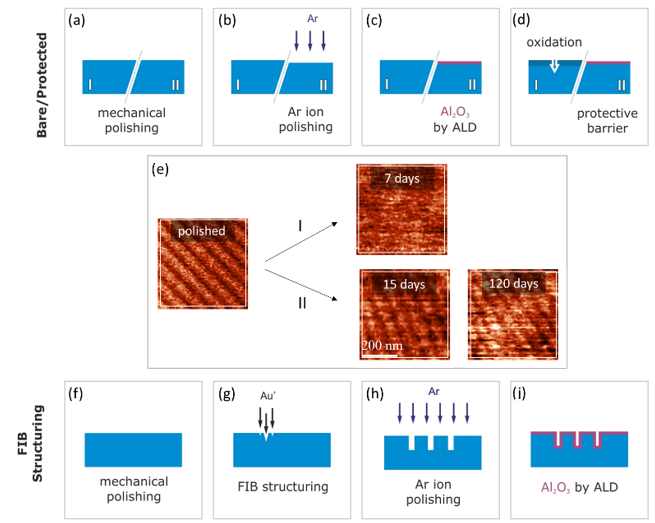}
    \caption{
        Sample preparation.
        (a)--(d) Preparation of FeGe single crystals for MFM.
        Panels (I) illustrate the conventional approach using only mechanical polishing (a), leading to rapid surface oxidation (d).
        The improved protocol (II) combines Ar ion polishing (b) with mechanical polishing (a), followed by deposition of a protective Al$_2$O$_3$ capping layer (c) via atomic layer deposition (ALD), which suppresses oxidation (d).
        (e) Without the capping layer (I), the MFM signal degrades within days, while the capped sample (II) retains a stable signal even after 120 days in air.
        (f)--(i) show the preparation sequence for structured FeGe surfaces protected by the Al$_2$O$_3$ layer.
    }
    \label{fig:S4}
\end{figure}

The preparation sequence and protective capping procedure summarized in Fig.~\ref{fig:S4} correspond to the FeGe lamellae studied in the main text.
The improved surface protection enabled stable long-term MFM measurements and prevented oxidation-related degradation of the helical spin texture.

\section*{Supplementary References}
\bibliographystyle{naturemag}
\bibliography{references}